\documentclass[a4paper,11pt]{article}
\usepackage{jheppub} 
\usepackage{jheppub} 
\usepackage{lineno}
\usepackage{float}

\usepackage{orcidlink}

\usepackage[font=small, labelfont=bf]{caption}
\usepackage{comment}
\usepackage[normalem]{ulem}
\usepackage[english]{babel}
\usepackage{graphicx}
\usepackage{dcolumn}
\usepackage{bm}
\usepackage{blindtext}
\usepackage{verbatim}
\usepackage{mathrsfs}
\usepackage{musicography}
\usepackage{amsmath}
\usepackage{dsfont}
\usepackage{mathrsfs}
\usepackage{amssymb}
\usepackage{amsthm}
\usepackage{cancel}
\usepackage{physics}
\usepackage{epstopdf}
\usepackage{mathtools}
\usepackage{color}
\usepackage[usenames,dvipsnames]{pstricks}
\usepackage{epsfig}
\usepackage{pst-grad} 
\usepackage{pst-plot} 
\usepackage{hyperref}
\usepackage{verbatim}
\usepackage{afterpage}
\usepackage{sidecap}
\usepackage{titlesec}
\titlespacing*{\subsection}{0pt}{1ex}{0.5ex}
\hypersetup{
    colorlinks=true,
    linkcolor=blue,
    filecolor=red,      
    urlcolor=cyan,
}
\usepackage{tikzsymbols}

\title{A novel quantum memory effect and thermal modulation in graviton-mediated entanglement}

\author[1]{Mainak Dutta \orcidlink{0009-0009-5265-6689},}
\author[2,3]{Partha Nandi \orcidlink{0000-0001-5290-494X},}
\author[1]{ Bibhas Ranjan Majhi \orcidlink{0000-0001-8621-1324}}
\affiliation[1]{Department of Physics,\\ Indian Institute of Technology Guwahati, Guwahati 781039, Assam, India}
\affiliation[2]{Department of Physics, \\University 
	of Stellenbosch, Stellenbosch-7600, South Africa}
\affiliation[3]{National Institute of Theoretical and Computational Sciences (NITheCS),\\Stellenbosch, 7604, South Africa}

\emailAdd{d.mainak@iitg.ac.in} \emailAdd{pnandi@sun.ac.za} \emailAdd{bibhas.majhi@iitg.ac.in}

\abstract{{A central challenge in probing the quantum nature of gravity is to distinguish effects that are genuinely quantum from those that can be explained classically. In this work, we study how quantized gravitational waves interact with thermal quantum systems, modeled as harmonic oscillators. We show that, unlike classical waves, quantized gravitons generate entanglement and leave behind a persistent “graviton-induced quantum memory” even after the wave has passed. This effect is further shaped by the presence of thermal noise, which does not simply wash out quantum correlations but can in fact amplify them in distinctive ways. Our analysis reveals clear signatures—such as nonlinear thermal corrections and a prethermal time-crystal–like phase—that cannot arise from any classical treatment. These results identify experimentally relevant markers of gravitons and provide a framework for exploring how finite-temperature environments may help uncover the quantum nature of gravity.}}

\begin{document}
\maketitle
\flushbottom


\section{Introduction}

Is gravity quantum? This question lies at the heart of modern physics, challenging our understanding of nature’s most fundamental force~\cite{DeWitt1967, Penrose1996, Kiefer2004}. The main obstacle lies in the inaccessibility of the Planck scale, where quantum gravity effects are expected to dominate~\cite{Nandi2024}. Yet, recent theoretical developments suggest that gravity’s quantum features may manifest indirectly, not through extreme energies, but via subtle quantum correlations between spatially separated matter systems~\cite{Das2008, Pikovski2012, vanDeKamp2020}. For example, two isolated objects could become entangled purely through their mutual gravitational interaction—a process impossible if gravity were classical. This realization has inspired experimental proposals, both tabletop and space-based, to test gravity’s quantum nature~\cite{Bose:2017nin}.  

Entanglement alone, however, does not fully capture gravity’s quantum fingerprint~\cite{Nandi:2024jyf}. A complete understanding requires a genuinely quantum description of gravity. In this work, we investigate how quantized gravitational waves (GWs) mediate correlations between realistic quantum detectors operating in thermal environments. Our motivation comes from the remarkable progress of laser interferometry, which has transformed astrophysics by enabling the direct detection of GWs~\cite{ LIGO2016}. While current detections predominantly involve transient signals from compact-object mergers, continuous or stochastic backgrounds are also expected from rotating neutron stars, unresolved astrophysical sources, and relic cosmological processes such as primordial black holes, cosmic strings, and the early universe~\cite{Carr:1974nx,Taylor:2019utq,Cruise:2012zz}. Additional stationary sources may include scalar-field dark matter~\cite{PhysRevResearch.1.033187} and quantum-gravity–induced stochastic fluctuations~\cite{PhysRevD.77.104031}. Modern interferometers such as LIGO~\cite{Hertzberg:2021rbl,PhysRevD.104.062006} and LISA, whose suspended test masses behave as harmonic oscillators approaching the quantum limit, therefore offer a promising platform to probe both classical and quantum aspects of GWs, potentially revealing signatures of gravity’s quantization~\cite{FrederikGScholtz:2025mcs}.  

Current large-scale detectors are most sensitive in the range \(10\ \mathrm{Hz}\)–\(10^3\ \mathrm{Hz}\), measuring strains of order \(h \sim 10^{-21}\), corresponding to mirror displacements of $10^{-18}\,\mathrm{m}$~\cite{Sigg1998,Nandi:2024jyf,Nandi:2024zxp}. For comparison, the quantum zero-point fluctuations of their \(40\,\mathrm{kg}\) mirrors are $\sim 10^{-20}\,\mathrm{m}$~\cite{Aston2012}. Such macroscopic test masses, however, decohere rapidly through environmental couplings, limiting their ability to display coherent quantum effects. This limitation has motivated the exploration of mesoscopic systems~\cite{OConnell2010,Teufel2011,Chan2011}, with masses \(10^{-12}\)–\(10^{-9}\,\mathrm{kg}\), which can be cooled close to their ground states~\cite{PhysRevLett.91.130401,Matsumoto2025}. In this regime, the zero-point uncertainty $\Delta x_{\mathrm{zpf}} \sim 10^{-17}\,\mathrm{m}$~\cite{Miki2024, Wan2017} becomes comparable to—or larger than—GW-induced displacements, bringing quantum and gravitational effects onto nearly equal footing.  

The conceptual groundwork for entanglement as a probe of quantum gravity was laid by Bose et al.~\cite{Bose:2017nin} and Marletto \& Vedral~\cite{Marletto:2017kzi}, who argued that two masses in spatial superposition could become entangled via Newtonian gravity. Detecting such gravity-mediated entanglement (GME) would provide strong evidence that gravity carries quantum degrees of freedom. While early proposals assumed static Newtonian interactions, subsequent extensions incorporated optomechanical and interferometric platforms~\cite{Carney:2018ofe,Howl:2023xtf,Christodoulou:2022mkf}. Our work advances this direction by fully quantizing the GW field, thereby revealing dynamical, frequency-dependent signatures of gravitons in experimentally realistic settings.  
Field quantization also predicts the graviton as a new quantum particle. Although Dyson argued that detecting a single graviton is practically impossible~\cite{Dyson:2013hbl}, later work showed that nonclassical states—such as squeezed or squeezed-thermal gravitons—could leave observable imprints~\cite{Parikh:2020nrd,PhysRevD.103.044017,PhysRevD.104.083516,Hsiang:2024qou}.  Understanding how these states influence quantum-sensitive detectors is therefore crucial. In particular, Manikandan and Wilczek \cite{Manikandan2025} have recently proposed concrete operational diagnostics — including counting statistics for resonant bar detectors and phase-sensitive (homodyne) readout strategies — that can distinguish coherent (classical-like) gravitational radiation from thermal or squeezed states.

The purpose of this paper is to develop a quantitative framework for analyzing how quantized gravitational waves (GWs) influence entanglement and the thermal dynamics of detectors. Specifically, we address the following questions: 
\textit{Can thermal noise enhance, rather than suppress, entanglement generation by quantum GWs? And are there measurable features in the thermal energy distribution that uniquely signal a quantum-gravitational origin?} 
These questions are crucial because no detector is perfectly isolated: thermal noise is unavoidable. Furthermore, realistic detectors need not share the same temperature as the ambient graviton background. Laboratory platforms—such as cryogenically cooled resonators—are designed to minimize noise, whereas a thermal graviton population, if it exists, would likely originate from cosmological relics or astrophysical processes. Allowing different detector and graviton temperatures, $ T $ and $ T' $, is therefore essential. This ``nonequilibrium'' setting captures stimulated graviton processes absent in equilibrium and reveals nonlinear thermal corrections that uniquely characterize quantum-gravitational interactions.


To investigate these effects, we model the detector as two decoupled one-dimensional harmonic oscillators—representing orthogonal vibrational modes of a mesoscopic mirror in an interferometric arm, such as Advanced LIGO—interacting with a thermal graviton environment. We employ the Thermo Field Dynamics (TFD) formalism to express thermal averages as quantum expectation values and use the Magnus expansion to derive perturbative dynamics upto second order. This approach connects with broader studies of environment-induced entanglement~\cite{Braun2002} and recent feasibility analyses of optomechanical and interferometric platforms~\cite{Carney2019pza,Carney2021ktp,Carney2021yfw}, while explicitly incorporating the graviton bath as a finite-temperature noise source. 

Our analysis yields closed-form expressions for the reduced purity, second-order Rényi entropy, and local occupation number in terms of universal kernel integrals over GW two-point correlators. From these expressions, two distinctive quantum signatures emerge:
\begin{itemize}
    \item \textbf{Graviton-induced quantum memory effect:} The two detector modes, initially uncoupled, develop correlated excitations due to their independent local interaction with the \emph{quantized plus-polarized} GW field. These excitations reflect entanglement generated between the modes even in the absence of direct coupling. Remarkably, the detector retains a persistent offset in its reduced density matrix even at a time when the GW drive is periodically ceased—a purely quantum effect absent in classical-wave treatments. In the zero-temperature graviton vacuum ($T' \to 0$), this memory persists and is accompanied by entanglement generation. Finite graviton temperatures ($T'$) introduce stimulated processes that combine with the detector temperature ($T$), producing nonlinear corrections up to cubic order in the detector occupation $\langle \hat{n}_\beta\rangle$, where $\beta = 1/T$. The distinct roles of $T$ and $T'$ reveal a genuinely new quantum–thermal interplay.


    \item \textbf{Prethermal time-crystal--like behavior:} If the detector’s internal frequency is detuned from the GW drive, its reduced density matrix exhibits stable subharmonic oscillations—effects absent in the classical dynamics of plus-polarized gravitational waves. Tracing over unobserved degrees of freedom—including gravitons and thermal partners—produces an emergent periodicity distinct from the period of the external quantized gravitational drive. Finite-temperature gravitons further enhance these oscillations through nonlinear corrections proportional to both $\langle \hat{n}_\beta\rangle$ and $\langle \hat{n}_{\beta'}\rangle$ with $\beta' = 1/T'$, revealing a nontrivial interplay between thermal noise and entanglement generation. While the quantum memory effect exists even without thermal effects, its interpretation as a prethermal time-crystal--like phase is meaningful only in a thermal environment, where the detector exhibits a graviton-induced nonlinear deformation of the Bose–Einstein distribution, reflecting higher-order thermal–quantum correlations.
\end{itemize}
This framework bridges theoretical proposals of gravity-mediated entanglement with experimental optomechanics, providing possible concrete, testable criteria for probing quantum-gravitational effects in finite-temperature environments.

The paper is structured as follows. Section~\ref{sec:H} introduces the system Hamiltonian describing two harmonic oscillators coupled to quantized gravitational wave modes, and outlines how classical cross-polarization effects are removed to isolate purely quantum correlations. Section~\ref{sec:TFD} applies the Thermo Field Dynamics (TFD) formalism to incorporate finite-temperature effects, constructing the initial thermal states of the detector system. Section~\ref{sec:Time-evolution} develops the perturbative time evolution of the thermal state using the Magnus expansion. Section~\ref{sec:entangle} analyzes the resulting entanglement phenomena, with subsections devoted to purity loss and the growth of entanglement entropy. Section~\ref{sec:therm-respons} focuses on the thermal response of the detector, identifying distinctive graviton-induced effects, including the emergence of a quantum memory offset and a prethermal time-crystal-like phase. Section~\ref{sec:conclu} concludes with a summary of the main findings and their implications for probing the quantum nature of gravity. Finally, the appendices provide technical details, including the derivation of the graviton Hamiltonian, the TFD construction of the thermal ground state, and explicit calculations of the evolved density matrix.

\section{System Hamiltonian and quantum gravitational interaction}\label{sec:H}

Gravitational waves (GWs) cause deviations in geodesics transverse to their propagation direction. At the linearized level, this effect can be effectively modeled by two particles confined to move independently along the \(x\) and \(y\) axes, which are perpendicular to the GW propagation along the \(z\)-axis. We consider initially each particle is trapped in its own harmonic potential along its respective direction, and the GW acts as a perturbation.
In the weak-field approximation, the dynamics under a GW traveling along the \(z\)-axis is governed by the time-dependent Hamiltonian \cite{Nandi:2022sjy}:
\begin{eqnarray}
\hat{H}(t) &=& \sum_{i=1}^{2} \left( \alpha \hat{P}_i^2 + \beta \hat{X}_i^2 \right)
+ \gamma(t)\left( \hat{X}_1 \hat{P}_1 + \hat{P}_1 \hat{X}_1 \right)
- \gamma(t)\left( \hat{X}_2 \hat{P}_2 + \hat{P}_2 \hat{X}_2 \right)
\nonumber
\\
&+& \delta(t)\left( \hat{X}_1 \hat{P}_2 + \hat{P}_1 \hat{X}_2 \right),
\label{eq:GW_Hamiltonian}
\end{eqnarray}
where 
$\alpha = 1/(2M), \quad \beta = (1/2) M \omega^2, \quad \gamma(t) = \frac{\dot{\chi}(t)}{2} \epsilon_+, \quad \delta(t) =  \dot{\chi}(t) \epsilon_\times$.
In the transverse-traceless gauge, the GW metric perturbation is 
\begin{equation}
h_{jk}(t) = 2\chi(t) \left( \epsilon_\times \sigma^1_{jk} + \epsilon_+ \sigma^3_{jk} \right),
\end{equation}
where \(\sigma^1\) and \(\sigma^3\) are Pauli matrices, and \(\epsilon_+\), \(\epsilon_\times\) denote plus and cross polarizations of GW. Here, \(2\chi(t)\) denotes the time-dependent GW amplitude. A detailed and systematic derivation of this Hamiltonian is given in \cite{PhysRevD.51.1701,Nandi:2024jyf} (also see the appendix of \cite{Dutta:2025bge}).
This model, initially introduced in \cite{PhysRevD.51.1701}, has been studied extensively in quantum gravity contexts \cite{Nandi:2022sjy,Nandi:2024zxp,Nandi:2024jyf}. Note that the cross polarization term \(\delta(t)\) induces classical coupling between the two oscillators, while the plus polarization terms act locally and independently on each oscillator.

To focus on correlations arising purely from quantum effects, we set the classical cross-polarization component $\varepsilon_\times = 0$. This can be done without loss of generality by applying a time-independent unitary transformation that diagonalizes the cross-polarization-induced coupling. Specifically, there exists a rotation in the $x,y$ mode space—generated by suitable linear combinations of $\hat{X}_i$ and $\hat{P}_i$—that maps the cross-coupling term into local quadratic terms. After performing this transformation and absorbing the resulting local shifts into the oscillator frequencies, one finds an effective Hamiltonian which is similar to  setting $\varepsilon_\times = 0$ in (\ref{eq:GW_Hamiltonian}) (see \cite{Nandi:2022sjy} for a detailed analysis). This effectively removes classical cross-coupling, allowing us to isolate correlations that arise purely from quantum interactions.
 The plus polarization terms then produce local, independent action on each oscillator, so the interaction picture time evolution operator factorizes as
\begin{equation}
\hat{U}^{\gamma}_{\mathrm{int}}(t) = \hat{U}^{\gamma}_{\mathrm{int},1}(t) \otimes \hat{U}^{\gamma}_{\mathrm{int},2}(t).
\label{UU}
\end{equation}
This factorization reflects the locality of the system and implies that entanglement cannot arise from the Hamiltonian alone unless the gravitational interaction involves inherently quantum degrees of freedom \cite{Nandi:2024jyf}.
When treated classically, GWs appear as time-dependent external fields modulating the system Hamiltonian, effectively prescribed classical drives that do not themselves evolve dynamically. However, a fully quantum treatment requires quantizing the GW field, introducing graviton creation and annihilation operators that dynamically couple to the system. This framework enables self-consistent evolution of the detector and quantum gravitational field, capturing vacuum fluctuations and entanglement beyond the classical description \cite{Nandi:2024jyf}.

For further analysis, we consider the interaction of the two harmonic oscillators with quantized gravitational wave (GW) modes. We define annihilation operators for the oscillators as  \begin{equation}
\hat{a}_i = \left(\frac{\alpha}{\beta}\right)^{1/4} \frac{\sqrt{\beta/\alpha} \, \hat{X}_i + i \hat{P}_i}{\sqrt{2\hbar}}, \quad i=1,2,
\end{equation}
which satisfy the canonical commutation relations
\begin{equation}
[\hat{a}_i, \hat{a}_j^\dagger] = \delta_{ij} \mathbb{I}, \quad i,j=1,2.
\end{equation}
On the other hand, in the interaction picture the gravitational interaction operator takes the form \cite{Nandi:2024jyf} 
\begin{equation}
\hat{\gamma}(t) = \frac{1}{2} \dot{\hat{\chi}}(t) = i C_\gamma \left( \hat{b}_g e^{-i \omega_g t} - \hat{b}_g^\dagger e^{i \omega_g t} \right),
\label{h}
\end{equation}
with coupling constant  
\begin{equation}
C_\gamma=-\sqrt{\frac{\omega_gc\pi l_p^2}{2L^3}}~,
\end{equation}
where \(\omega_g\) is the GW frequency, \(L\) the quantization box size, and \(l_p\) the Planck length.
Then the total Hamiltonian (\ref{eq:GW_Hamiltonian}) for plus polarization, including the quantized graviton mode, can then be expressed as
\begin{equation}
\hat{H}(t) = \mathbb{I}_{D}\otimes\underbrace{\hbar \omega_g \left( \hat{b}_g^\dagger \hat{b}_g + \frac{1}{2} \right)}_{\hat{H}_G}
+ \underbrace{2 \hbar \sqrt{\alpha \beta} \left( \sum_{i=1}^2 \hat{N}_i + 1 \right)}_{\hat{H}_D}\otimes\mathbb{I}_{G}
+ \underbrace{i \hbar \left( \hat{a}_1^{\dagger 2} - \hat{a}_1^2 - \hat{a}_2^{\dagger 2} + \hat{a}_2^2 \right) \otimes \hat{\gamma}(t)}_{\hat{H}_{\mathrm{int}}(t)},
\end{equation}
where \(\hat{N}_i = \hat{a}_i^\dagger \hat{a}_i\), \(\hat{b}_g\) denotes the graviton annihilation operator associated with plus-polarized modes, and \(\hat{\gamma}(t)\) is the time-dependent GW interaction operator. A detailed discussion of \(\hat{H}_G\) is provided in the Appendix \ref{Appendix A}.

It is important to emphasize that, for simplicity, we consider only a single quantized mode of the gravitational field interacting with the detector system. This single-mode approximation captures the essential quantum features—such as vacuum and thermal fluctuations, as well as mediator-induced correlations—while keeping the perturbative calculations tractable. The approximation is physically well-motivated for several reasons:

(i) In the presence of a nearly monochromatic GW source, such as continuous waves from a rotating neutron star, the detector response is dominated by a single spectral line.  

(ii) More generally, in the interaction picture, the detector–graviton interaction Hamiltonian is linear in the graviton operators,
\begin{equation}
\hat{H}_\mathrm{int}^I(t) = e^{\frac{i}{\hbar}\hat{H}_{D}t} \hat{H}_{int}(t)e^{-\frac{i}{\hbar}\hat{H}_{D}t}~. 
\end{equation}
In this case we consider an initial state
\begin{equation}
|i\rangle = |n_1, n_2; 0_g\rangle,
\end{equation}
with detector occupation numbers \(n_1, n_2\) and the graviton mode in the vacuum, and a final state
\begin{equation}
|f\rangle = |n_1+2, n_2; 1_g\rangle,
\end{equation}
corresponding to the absorption of a single graviton by the first oscillator. The first-order transition amplitude is then
\begin{equation}
\mathcal{A}_{i\to f}(T) = \frac{1}{i \hbar} \int_0^T dt \, \langle f | \hat{H}_\mathrm{int}^I(t) | i \rangle 
\sim \int_0^T dt \, e^{i (\omega_0 - \omega_g) t},
\end{equation}
where \(\omega_0 = 2\omega\) is the detector transition frequency. Rapid oscillations suppress off-resonant graviton modes ($|\omega_0 - \omega_g| \gg 1/T$), so only modes near resonance contribute significantly. Hence, the single-mode approximation captures the dominant dynamics of the detector–graviton system.  


\medskip

We further note that, at the initial time \(t=0\), the composite state can be written as
\begin{equation}
|\Psi(0)\rangle = (|0\rangle_1 \otimes |0\rangle_2)_D \otimes |0_g\rangle \equiv |00;0_g\rangle_{t=0}, 
\end{equation}
with both oscillators in their ground states and the graviton field in the vacuum. Although the interaction-picture time-evolution operator factorizes as in Eq.~(\ref{UU}), this factorization does not imply that the final state remains separable. Each operator \(\hat{U}^\gamma_{\mathrm{int},i}(t)\) depends on the common graviton mode \(\hat{\gamma}^I(t)\), thereby coupling the oscillators indirectly. As a result, initially separable states evolve into entangled, non-separable states via their mutual interaction with the quantized gravitational field, as studied in great detail in \cite{Nandi:2024jyf}.

Importantly, while the global evolution is unitary and preserves the purity of the total state, tracing out the gravitational degrees of freedom leaves the detectors in a mixed and generally entangled state. This highlights that quantum gravitational fluctuations can mediate correlations between otherwise independent subsystems—a feature entirely absent in a classical GW description, where no entanglement arises from initially factorized states.

\section{Thermalization via Thermofield Dynamics} \label{sec:TFD}

Following the program outlined above, we construct the initial state of the two-mode HO system such that the detector oscillators are prepared in thermal states. To incorporate thermal effects, we employ the \textit{Thermofield Dynamics} (TFD) formalism~\cite{Das1997,TakahashiUmezawa1996,BiswasDas1988}, which recasts a thermal mixed state as a pure state in an enlarged Hilbert space by introducing auxiliary (tilde) degrees of freedom. In this formulation, the detector’s thermal state appears as a pure state of the extended system, rather than as an entangled state of the physical oscillators themselves. The TFD construction is applied solely to the detector sector, while the graviton field is kept in a general Fock-space state. This separation ensures that the detector’s intrinsic thermal statistics are cleanly distinguished from the quantum correlations generated by its coupling to the quantized gravitational field.


The total Hamiltonian in this formalism is updated as
\begin{equation}
\hat{\mathcal{H}}(t) = \hat{\mathcal{H}}_0 + \hat{\mathcal{H}}_{\mathrm{int}}(t),
\end{equation}
where
\begin{align}
\hat{\mathcal{H}}_0 &=\mathbb{I}_D \otimes \hbar \omega_g \hat{b}_g^\dagger \hat{b}_g +\sum_{i=1}^2 \hbar \omega \left( \hat{a}_i^\dagger \hat{a}_i - \hat{\tilde{a}}_i^\dagger \hat{\tilde{a}}_i \right) \otimes \mathbb{I}_G  , \label{eq:H0_TFD}\\[6pt]
\hat{\mathcal{H}}_{\mathrm{int}}(t) &= i \hbar \Big\{ \left[ \left( \hat{a}_1^{\dagger 2} - \hat{a}_1^2 \right) \otimes \mathbb{I}_2 - \mathbb{I}_1 \otimes \left( \hat{a}_2^{\dagger 2} - \hat{a}_2^2 \right) \right] \notag \\
&\quad - \left[ \left( \hat{\tilde{a}}_1^{\dagger 2} - \hat{\tilde{a}}_1^2 \right) \otimes \mathbb{I}_{\tilde{2}} - \mathbb{I}_{\tilde{1}} \otimes \left( \hat{\tilde{a}}_2^{\dagger 2} - \hat{\tilde{a}}_2^2 \right) \right] \Big\} \otimes \hat{\gamma}(t).
\label{eq:Hint_TFD} 
\end{align}
Here, it may be noted that the ground-state energy (i.e. $(1/2)\hbar \omega_g$ factor) contribution has been omitted, since it merely produces a uniform constant shift to all energy levels. Particularly when we will find the average value of any observable, then it will not contribute as it cancels from the normalization of density matrix. Therefore in our subsequent computations within the interaction picture, as this term plays no significant role, we exclude this from the Hamiltonian. 
At the initial time, the thermal vacuum state of each harmonic oscillator (HO) is given by \cite{Das1997}  
\begin{equation}
\ket{0; t=0}_\beta = (1 - e^{-\beta\hbar \omega})^{1/2} 
\sum_{n=0}^\infty e^{-\frac{\beta\omega\hbar n}{2}} \ket{n,\tilde{n}}~,
\label{B1}
\end{equation}
where \(\beta = 1/T\). In fact for brevity we keep $\hbar=k_{B}=c=1$ for our computation. Tracing out the tilde degrees of freedom in the corresponding density matrix reproduces the expected thermal state at inverse temperature \(\beta\).   Consequently, at \(t=0\), for two coupled oscillators and a generic graviton state \(\ket{G}\) (not necessarily the vacuum), the initial thermal state reads 

\begin{eqnarray}
\ket{0,0;G; t=0}_\beta 
&=& \ket{0; t=0}_\beta^{(1)} \otimes \ket{0; t=0}_\beta^{(2)} \otimes \ket{G}
\nonumber
\\
&=& \Big(1 - e^{-\beta \omega}\Big) \sum_{n_x,n_y=0}^\infty e^{-\frac{\beta \omega}{2} (n_x + n_y)} 
\ket{n_x, n_y, \tilde{n}_x, \tilde{n}_y; G}.
\label{thermal_gs_t0}
\end{eqnarray}
Here, the total Hilbert space is the doubled space \(\mathcal{H} = \mathbf{H} \otimes \tilde{\mathbf{H}}\), spanned by states  
$\ket{n_1,n_2; \tilde{m}_1, \tilde{m}_2}_{D} = \ket{n_1,n_2} \otimes \widetilde{\ket{m_1,m_2}}$,  
with $\ket{n_1,n_2}$ and $\widetilde{\ket{m_1,m_2}}$ denoting eigenstates of the physical and tilde copies of a two-dimensional isotropic oscillator of frequency \(\omega\).  
Although the thermal vacuum \eqref{thermal_gs_t0} is a pure state in the extended Hilbert space \(\mathcal{H}\), its restriction to the physical sector yields a mixed state. Explicitly,  
\begin{equation}
\hat{\rho}^\beta = \mathrm{Tr}_{\tilde{\mathbf{H}}} \Big( \ket{0,0;G;t=0}{_\beta}~{_\beta}\bra{0,0;G;t=0} \Big),
\end{equation}
reproduces the standard Gibbs equilibrium thermal state. This essential feature enables Thermo Field Dynamics (TFD) to provide a fully unitary description of finite-temperature quantum systems within the operator formalism: the total system evolves under \(\hat{\mathcal{H}}(t)\), while the physical sector accounts for thermal fluctuations through the partial trace over the tilde space.




This formalism is particularly well suited to our setup because two distinct sources of correlations coexist:
\begin{itemize}
    \item The GW induces coherent, time-dependent two-mode squeezing interactions that entangle the two physical oscillators,
    \item The thermal environment manifests as initial entanglement between each physical oscillator and its corresponding fictitious thermal partner introduced in the TFD construction.
\end{itemize}
The strength of the TFD framework is that it allows these two channels of entanglement to be disentangled algebraically and tracked within a unified operator formalism. This makes it possible to identify precisely how temperature modifies—but does not independently generate—the gravitationally induced entanglement. In this sense, temperature acts as a \emph{catalyst}, amplifying the correlations seeded by the spacetime geometry, a feature that is obscured in conventional mixed-state thermal approaches.

It is important to emphasize that at the initial time $t=0$, the detector thermal vacua and the graviton sector are not entangled. Moreover, the two detector oscillators and their respective tilde partners are also not coupled to each other; each thermal vacuum (see Eq.~(\ref{B1})) represents correlations only between a given oscillator and its own fictitious tilde mode. Consequently, if one traces out the second oscillator together with its tilde partner and the graviton sector, the reduced state of the first oscillator is simply a thermal Gibbs state. Similarly, tracing out the tilde mode of the first oscillator reproduces the familiar mixed-state thermal description. 
However, once the state is evolved with respect to the full interaction Hamiltonian in the interaction picture, new channels of entanglement emerge: between the two oscillators, between detector and graviton modes, and across tilde sectors. The central objective of our analysis is to isolate the genuinely gravitationally induced entanglement from this structure and to pinpoint its dependence on temperature and memory effects.


Finally, for the graviton field we initially assume a general pure Fock-space state. Thermal effects can be incorporated later at the level of expectation values of graviton creation and annihilation operators, either by adopting the TFD formalism for the graviton sector or, equivalently, by ensemble averaging over Gibbs states. Since only correlators of graviton operators enter our analysis, both approaches are equivalent and the choice does not affect the results. For definiteness, we therefore keep the graviton sector as a pure state, while noting that a thermal description could be adopted without altering our conclusions.
\\
\section{Time evolution of initial state}\label{sec:Time-evolution}
Due to the time dependence in the interaction Hamiltonian, we work in the interaction picture and treat the coupling perturbatively, assuming a weak gravitational interaction. In this picture the time-evolved state is
\begin{equation}\label{t-evolved-state}
\ket{0,0;G}^I_{\beta,t} = \hat{U}^\gamma_{\mathrm{int}}(t) \ket{0,0;G}_{\beta(t=0)},
\end{equation}
where the interaction-picture time-evolution operator is
\begin{equation}
\hat{U}^\gamma_{\mathrm{int}}(t) = \mathrm{\bf T} \Big[\exp \left( -i \int_0^t \hat{\mathcal{H}}_{\mathrm{int}}^I(t') dt' \right)\Big],
\label{eq:unitary_operator}
\end{equation}
with \(\mathrm{\bf T}\) denoting time ordering, and
\begin{equation}
\hat{\mathcal{H}}_{\mathrm{int}}^I(t) = e^{i \hat{\mathcal{H}}_0 t} \hat{\mathcal{H}}_{\mathrm{int}}(t) e^{-i \hat{\mathcal{H}}_0 t}.
\label{eq:interaction_hamiltonian}
\end{equation}
Employing the Magnus expansion~\cite{Magnus1954}, the unitary operator can be expanded perturbatively up to second order in the gravitational interaction operator \(\hat{\gamma}(t)\):
\begin{eqnarray}\label{U_Magnus}
\hat{U}^\gamma_{\mathrm{int}}(t) &\approx& 1
- i \int_0^t \hat{H}_{\mathrm{int}}(t_1) dt_1 
+ i \int_0^t \hat{\widetilde{H}}_{\mathrm{int}}(t_1) dt_1 
- \frac{1}{2} \int_0^t dt_1 \int_0^{t_1} dt_2 [ \hat{H}_{\mathrm{int}}^I(t_1), \hat{H}_{\mathrm{int}}^I(t_2) ] 
\nonumber
\\
&-& \frac{1}{2} \int_0^t dt_1 \int_0^{t_1} dt_2 [ \hat{\widetilde{H}}_{\mathrm{int}}^I(t_1), \hat{\widetilde{H}}_{\mathrm{int}}^I(t_2) ] 
- \frac{1}{2} \left( \int_0^t \left( \hat{H}_{\mathrm{int}}^I(t_1) - \hat{\widetilde{H}}_{\mathrm{int}}^I(t_1) \right) dt_1 \right)^2.
\end{eqnarray}
In the interaction picture, the evolved thermal vacuum state of the composite oscillator detector--graviton system, truncated at second order in the interaction strength, can be written compactly as
\begin{eqnarray}
\label{eq:thermal_state_compact}
&&\ket{0,0;G;t}_{\beta}^I =
\left( 1 - e^{-\beta \omega} \right)
\sum_{n_x, n_y \ge 0} e^{ -\frac{\beta\omega}{2} (n_x + n_y) }
\sum_{k=0}^{2} \;
\sum_{(a,b,c,d) \in S_k}
(\mathbb{I_{D}}\otimes \hat{C}^{(k)}_{abcd}(t)) \;
\nonumber
\\
&&\ket{n_x+a,\; n_y+b,\; \tilde{n}_x+c,\; \tilde{n}_y+d}\otimes \ket{G} .
\nonumber
\\
\end{eqnarray}
Here, $\mathbb{I}_{D}=(\mathbb{I}_{1}\otimes\mathbb{I}_{\tilde{1}})\otimes(\mathbb{I}_{2}\otimes\mathbb{I}_{\tilde{2}})$, and the set $S_k$ denotes the occupation-shift tuples arising at perturbative order $k$ (a derivation is presented in Appendix \ref{Appendix C}). The relevant sets for the present calculation, which reproduce the explicit long-form expansion of Eq.~(\ref{eq:thermal_state_compact}), are
\begin{align}
S_0 &= \{ (0,0,0,0) \} \, , \\
S_1 &= \{ (2,0,0,0),\ (0,2,0,0),\ (0,0,2,0),\ (0,0,0,2) \} \, , \\
S_2 &= \{ (0,0,0,0),\ (4,0,0,0),\ (0,4,0,0),\ (0,0,4,0),\ (0,0,0,4), \nonumber\\
&\quad (2,2,0,0),\ (0,0,2,2),\ (0,2,2,0),\ (2,0,0,2),\ (0,2,0,2),\ (2,0,2,0) \} \, .
\end{align}
It is important to emphasize that the perturbative order $k$ is \emph{independent} of the total occupation shift $a+b+c+d$. In particular, amplitude corrections that do not change the occupation numbers, such as $\hat{C}^{(2)}_{0000}$, are included via $(0,0,0,0) \in S_2$,  and \( \hat{C}^{(0)}_{0000} = 1 \). The nonzero coefficients up to second order are:
\begin{align}
\hat{C}^{(1)}_{2000} &= \alpha^{(1)}_1, \quad \hat{C}^{(1)}_{0200} = \alpha^{(1)}_3, \quad \hat{C}^{(1)}_{0020} = \alpha^{(1)}_2, \quad \hat{C}^{(1)}_{0002} = \alpha^{(1)}_4, \notag \\
\hat{C}^{(2)}_{0000} &= \delta^{(2)} + \epsilon^{(2)}_2 + \beta^{(2)}_{5,6} + \xi^{(2)}_{5,6}, \notag \\
\hat{C}^{(2)}_{4000} &= \hat{C}^{(2)}_{0040}=\beta^{(2)}_1 + \lambda^{(2)}_1 + \xi^{(2)}_1,  \notag \\
\quad 
\hat{C}^{(2)}_{0400} &= \hat{C}^{(2)}_{0004}=\beta^{(2)}_2 + \lambda^{(2)}_2 + \xi^{(2)}_2,  \notag \\
\hat{C}^{(2)}_{2200} &= \hat{C}^{(2)}_{0022} = \beta^{(2)}_3 + \lambda^{(2)}_3 + \xi^{(2)}_3, \notag \\
\hat{C}^{(2)}_{0220} &= \beta^{(2)}_4 + \gamma^{(2)}_1 + \epsilon^{(2)}_1 + \xi^{(2)}_4, \notag\\ 
\hat{C}^{(2)}_{2002} &= \beta^{(2)}_4 + \gamma^{(2)}_1 + \epsilon^{(2)}_3 + \xi^{(2)}_4, \notag \\
\hat{C}^{(2)}_{0202} &= \gamma^{(2)}_2, \quad \hat{C}^{(2)}_{2020} = \gamma^{(2)}_3~;
\label{Int_pic_coeff1}
\end{align}
where
the time-dependent integrals defining these coefficients are:
\begin{align}
\delta^{(2)} &= 8i(n_x + n_y + 1) \!\int_0^t\! \!\!dt_1 \int_0^{t_1} \!\!\!dt_2\, \hat{\gamma}^I(t_1)\hat{\gamma}^I(t_2)\sin\big(2\omega(t_1 - t_2)\big), \notag \\
\alpha^{(1)}_r &= \int_0^t \!\!dt_1\, \hat{\gamma}^I(t_1)\left( e^{2i\omega t_1} + e^{-\beta\omega} e^{-2i\omega t_1} \right) E^r, \notag \\
\beta^{(2)}_j &= \frac{1}{2} \!\int_0^t \!\!dt_1 \!\int_0^{t_1} \!\!dt_2\, [\hat{\gamma}^I(t_1), \hat{\gamma}^I(t_2)]\, d_j, \quad
\xi^{(2)}_j = \frac{1}{2} \int_0^t\! \!\!dt_1 \int_0^t \!\!\!dt_2\, \hat{\gamma}^I(t_2)\hat{\gamma}^I(t_1) d_j, \notag \\
\lambda^{(2)}_l &= \int_0^t\! \!\!dt_1 \int_0^t\! \!\!dt_2\, \hat{\gamma}^I(t_2)\hat{\gamma}^I(t_1) c_l, \quad
\epsilon^{(2)}_k = \int_0^t\! \!\!dt_1 \int_0^t\! \!\!dt_2\, \hat{\gamma}^I(t_2)\hat{\gamma}^I(t_1) b_k, \notag \\
\gamma^{(2)}_i &= \int_0^t\! \!\!dt_1 \int_0^t\! \!\!dt_2\, \hat{\gamma}^I(t_2)\hat{\gamma}^I(t_1) a_i.
\label{coeff0}
\end{align}
Here, \( E^q = (-1)^{q-1} \sqrt{(n_x + 2)(n_x + 1)} \), \( E^{q+2} = (-1)^{q} \sqrt{(n_y + 2)(n_y + 1)} \), with index ranges: \( i = 1,\dots,3 \); \( j = 1,\dots,6 \); \( k = 1,2,3 \); \( l = 1,\dots,3 \) and $q=1,2$. The mode functions \( a_i, b_k, c_l, d_j \) are defined in Eq.~\eqref{abcd}.

Our first step is to compute the density matrix derived from the time-evolved thermal ground state (g.s.). In the next section, we will explicitly evaluate entanglement measures such as purity and entanglement entropy, and then discuss the presence of gravity-induced entanglement between the modes of a single harmonic oscillator (HO).
We begin by tracing out the gravitational degrees of freedom:
\begin{equation}
    \hat{\rho}^\beta(t) = \mathrm{Tr}_G\big(\hat{\rho}^\beta_f(t)\big) = \sum_{n=0}^\infty \langle n_G | \hat{\rho}^\beta_f(t) | n_G \rangle,
    \label{rho}
\end{equation}
where $\hat{\rho}^\beta_f(t) = \ket{0,0;G;t}^{I}_{\beta} \,^{I}_{\beta}\bra{0,0;G;t}$
is the full density matrix including gravitational modes (denoted by the subscript \( f \)).
Tracing over these gravitational degrees of freedom yields a reduced description of the system that incorporates the influence of the quantum gravitational environment.  


\section{Entanglement phenomenon}\label{sec:entangle} 

We now turn to the qualitative analysis of the entangled nature of the final state. 
In the TFD framework, the thermal vacuum is initially a pure state, where each 
physical oscillator is entangled only with its corresponding fictitious (tilde) mode. 
At this stage, different physical oscillators remain uncorrelated, and tracing out 
the tilde modes simply reproduces the standard thermal mixed state for each 
oscillator.  

Once the quantized gravitational field is switched on, however, the situation changes: 
the two oscillators become indirectly coupled through their interaction with the common 
graviton environment. This generates additional correlation channels beyond the intrinsic 
thermal ones. In particular:  
\begin{enumerate}
    \item \textbf{Thermal entanglement:} Each physical oscillator remains entangled 
    with its tilde partner, as required in TFD.  
    \item \textbf{Graviton-mediated correlations:} The shared gravitational modes 
    induce effective couplings between the two otherwise independent physical oscillators.  
    \item \textbf{Cross-channel correlations:} Through the graviton-mediated interaction, 
    the tilde mode of one oscillator can become indirectly correlated with the physical 
    or tilde mode of the other.  
\end{enumerate}  
As a result, the time-evolved state of the system in the interaction picture can no 
longer be factorized into independent oscillator, tilde, and graviton sectors. After 
tracing out the graviton degrees of freedom, the reduced density matrix of the 
oscillators exhibits genuine entanglement generated dynamically by the gravitational 
interaction. This provides the qualitative foundation for a more detailed, quantitative 
analysis of the evolved thermal state.  
Our interest, therefore, lies in quantifying the entanglement generated between two 
harmonic oscillators (HOs) that are initially noninteracting at the classical level, 
with correlations arising exclusively through their indirect coupling to gravitational 
waves (GWs). To probe this, we consider the evolved thermal state of the first 
oscillator and apply appropriate entanglement measures.  

The first step involves tracing out the degrees of freedom of the second oscillator, including both its physical and fictitious (tilde) modes, thereby obtaining the reduced density matrix for the first oscillator and its associated tilde modes:  
\begin{align}\label{thermal_density_matrix_final_compact}
    &\hat{\rho}^\beta_{1\tilde{1}}(t) =\mathrm{Tr}_{2,\tilde{2}}\big(\hat{\rho}^\beta(t)\big)=\sum_{n_x,n'_x =0}^{\infty} \Bigg[
    \left(A^{(0)}_{n_x,n'_x} + B^{(2)}_{n_x,n'_x}\right) \ket{n_x,\tilde{n}_x}\bra{n'_x,\tilde{n}'_x} \notag \\
    &\quad + \biggl(
        D^{(1)}_{n_x,n'_x} \ket{n_x,\tilde{n}_x+2}
        + E^{(1)}_{n_x,n'_x} \ket{n_x+2,\tilde{n}_x}
        + H^{(2)}_{n_x,n'_x} \ket{n_x+4,\tilde{n}_x}
        + H^{(2)}_{n_x,n'_x} \ket{n_x,\tilde{n}_x+4}
    \biggr) \bra{n'_x,\tilde{n}'_x} \notag \\
    &\quad + \ket{n_x,\tilde{n}_x} \biggl(
        D'^{\dagger(1)}_{n_x,n'_x} \bra{n'_x,\tilde{n}'_x+2}
        + E'^{\dagger(1)}_{n_x,n'_x} \bra{n'_x+2,\tilde{n}'_x}
        + H'^{\dagger(2)}_{n_x,n'_x} \bra{n'_x+4,\tilde{n}'_x}
        + H'^{\dagger(2)}_{n_x,n'_x} \bra{n'_x,\tilde{n}'_x+4}
    \biggr) \notag \\
    &\quad + F^{(2)}_{n_x,n'_x} \biggl(
        \ket{n_x,\tilde{n}_x+2}\bra{n'_x,\tilde{n}'_x+2}
        + \ket{n_x+2,\tilde{n}_x}\bra{n'_x+2,\tilde{n}'_x}
    \biggr) \notag \\
    &\quad + G^{(2)}_{n_x,n'_x} \biggl(
        \ket{n_x+2,\tilde{n}_x}\bra{n'_x,\tilde{n}'_x+2}
        + \ket{n_x,\tilde{n}_x+2}\bra{n'_x+2,\tilde{n}'_x}
    \biggr)
    \Bigg].
\end{align}  
Here, terms are retained up to second order in the gravitational wave coupling \(\gamma\). The explicit forms and derivations of these coefficients are provided in Appendix \ref{Appendix D}.  
To further isolate the physical first oscillator, we trace out its corresponding fictitious tilde mode to obtain the reduced density matrix  
\begin{equation}
    \hat{\rho}^\beta_1(t) = \mathrm{Tr}_{\tilde{1}} \hat{\rho}^\beta_{1\tilde{1}} = \hat{\rho}^\beta_1(0) + \delta \hat{\rho}^\beta_g(t),
    \label{reduced_density_matrix}
\end{equation}  
where  
\begin{equation}
    \hat{\rho}^\beta_1(0) = \sum_{n=0}^\infty A^{(0)}_{n,n} \ket{n}\bra{n}~,
\end{equation}
corresponds to the initial thermal density matrix of the first oscillator, describing a mixed state. In contrast, the unperturbed part of \eqref{thermal_density_matrix_final_compact} corresponds to a pure state and so \eqref{thermal_density_matrix_final_compact} represents the evolved thermal vacuum state. Therefore, \eqref{reduced_density_matrix} represents the evolved thermal state of the first oscillator, including corrections due to gravitational interaction:  
\begin{equation}
\resizebox{\textwidth}{!}{$
\begin{aligned}
\delta \hat{\rho}^\beta_g(t) &= \sum_{n=0}^\infty \Big[ 
B^{(2)}_{n,n} \ket{n}\bra{n} 
+ D^{(1)}_{n,n+2} \ket{n}\bra{n+2} 
+ E^{(1)}_{n,n} \ket{n+2}\bra{n} 
+ D'^{\dagger(1)}_{n+2,n} \ket{n+2}\bra{n} \\
&\quad + E'^{\dagger(1)}_{n,n} \ket{n}\bra{n+2} 
+ F^{(2)}_{n,n} \ket{n}\bra{n} 
+ G^{(2)}_{n+2,n} \ket{n+4}\bra{n} 
+ G^{(2)}_{n,n+2} \ket{n}\bra{n+4} \\
&\quad + F^{(2)}_{n,n} \ket{n+2}\bra{n+2} 
+ H^{(2)}_{n,n} \ket{n+4}\bra{n} 
+ H^{(2)}_{n,n+4} \ket{n}\bra{n+4} \\
&\quad + H'^{\dagger(2)}_{n,n} \ket{n}\bra{n+4} 
+ H'^{\dagger(2)}_{n+4,n} \ket{n+4}\bra{n} 
\Big].
\end{aligned}
$}
\end{equation}  
The coefficients here are defined in Appendix \ref{Appendix D}. Notably, in the zero-temperature limit \(n=0, \beta \to \infty\), this expression reduces to the result obtained in \cite{Nandi:2024jyf}. 

At this stage, a few comments are in order. First, the reduced density operator 
\(\hat{\rho}^\beta_{1\tilde{1}}(t)\) is obtained by tracing out the graviton modes as well as 
the second physical oscillator and its corresponding fictitious (tilde) mode. This captures 
the mixedness arising solely from the indirect gravitational coupling between the two 
physical oscillators and their respective tilde partners. Subsequently, by further tracing 
out all fictitious modes, we obtain 
\(\hat{\rho}^\beta_1(t)\), which includes mixedness originating both from the indirect 
gravitational coupling and the intrinsic thermal correlations between each physical 
oscillator and its tilde mode. In the next subsection, we will show how these two types 
of reduced density matrices play a crucial role in quantifying the generated entanglement 
in terms of purity and entanglement entropy.

\subsection{Loss of purity}  \label{en}
To quantify the entanglement generated by gravitational interactions, we examine the purity, defined as \(P = \mathrm{Tr}(\hat{\rho}^2)\) for a state denoted by \(\hat{\rho}\). It is useful to introduce distinct notions of purity associated with the reduced density matrix, distinguishing intermediate and final (physical) forms:  
\begin{eqnarray*}
&&\hat{\rho}^{\beta}_\text{f} \xrightarrow{\text{trace over 2nd physical + 2nd fictitious modes+graviton}} P_Q^\beta(t)~,
\nonumber
\\
&&\text{(intermediate, dependent on the 1st fictitious)}
\end{eqnarray*}
and
\begin{eqnarray*}
&&\hat{\rho}^{\beta}_\text{f} \xrightarrow{\text{trace over 2nd physical + all fictitious + graviton}} P_{1Q}^\beta(t)~.
\nonumber
\\
&& \text{(physical, representing the first oscillator)}
\end{eqnarray*}

Starting from the reduced density matrix (\ref{thermal_density_matrix_final_compact}) and keeping terms up to \(\mathcal{O}(\gamma^2)\), the purity of the evolved state is found to be  
\begin{align}\label{purity_explicit}
    P^{\beta}_{Q}(t)  
        =& 1-16C_\gamma^2\left\{\frac{e^{-2\beta\omega}}{(1 - e^{-\beta\omega})^2}A(t)+\frac{2}{(1 - e^{-\beta\omega})^2}B(t)+\frac{e^{-\beta\omega}}{(1 - e^{-\beta\omega})^2}2F(t)\right\}\notag\\
        =&1-16C_\gamma^2~\chi_{\beta'}(\beta)
\end{align}  
where 
\begin{eqnarray}
&&A(t) = 2i \displaystyle \int_0^t dt_1 \int_0^{t_1} dt_2 \langle [\hat{\gamma}^I(t_1), \hat{\gamma}^I(t_2)] \rangle_G \sin T, 
\nonumber
\\
&&B(t) = \displaystyle \int_0^t dt_2 \int_0^t dt_1 \langle \hat{\gamma}^I(t_2) \hat{\gamma}^I(t_1) \rangle_G e^{T_-},
\nonumber
\\
&& F(t) = \displaystyle \int_0^t dt_2 \int_0^t dt_1 \langle \hat{\gamma}^I(t_2) \hat{\gamma}^I(t_1) \rangle_G \cos[2 \omega (t_1 + t_2)]~.
\label{BB1}
\end{eqnarray}
Here the subscript \(Q\) indicates that the purity loss arises from quantum gravitational wave effects.  
As expected, in the absence of coupling (\(\hat{\gamma}(t) = 0\)) the purity reduces to unity, confirming that temperature alone does not produce entanglement: the thermal vacuum state \(\hat{\rho}^\beta_{1\tilde{1}}(t)\) remains pure.  
Although the precise magnitude of entanglement depends on the detailed evaluation of the coefficients 
$\alpha^{(1)}_i, \beta^{(2)}_j, \delta^{(2)}, \xi^{(2)}_j, \epsilon^{(2)}_k, \gamma^{(2)}_i$, 
Eq.~\eqref{purity_explicit} clearly demonstrates that the \emph{order of entanglement} arises from quantum 
fluctuations of the gravitational field. Each physical mode is thermally coupled to a fictitious partner; 
tracing out the second physical oscillator along with its fictitious counterpart removes correlations, 
rendering the reduced state mixed. Consequently, the resulting impurity reflects entanglement mediated 
by graviton fluctuations that is no longer accessible after tracing.

To isolate gravitational contributions from purely thermal effects, we compute the purity of a single 
physical oscillator by tracing out only the fictitious modes from (\ref{reduced_density_matrix}):  
\begin{align}\label{purity_1Q}
P^\beta_{1Q}(t) &= \mathrm{Tr}_1 \left[ (\hat{\rho}^\beta_1(t))^2 \right] \notag \\
&= \left(\frac{1 - e^{-\beta \omega}}{1 + e^{-\beta \omega}}\right) - \frac{2}{(1 + e^{-\beta \omega})^2 (1 - e^{-\beta \omega})} \Big[ 2(1 + e^{-3 \beta \omega})(1 - e^{-\beta \omega}) + e^{-5 \beta \omega} (2 - e^{-\beta \omega}) \Big] B(t) \notag \\
&\quad + \frac{4 e^{-2 \beta \omega}}{(1 + e^{-\beta \omega})^2} C(t) - \frac{2 e^{-6 \beta \omega}}{(1 + e^{-\beta \omega})^3 (1 - e^{-\beta \omega})} \Big[ (2 - e^{-\beta \omega}) - e^{-\beta \omega} (1 - e^{-\beta \omega})^2 \Big] C^\dagger(t) \notag \\
&\quad + \frac{4 e^{-3 \beta \omega}}{(1 + e^{-\beta \omega})^3} (2 - e^{-4 \beta \omega}) F(t) - 2 e^{-\beta \omega} \frac{(2 - e^{-4 \beta \omega})}{(1 + e^{-\beta \omega})^2} \left(\frac{1 + e^{-2 \beta \omega}}{1 - e^{-2 \beta \omega}} \right) X(t) \notag \\
&\quad + 2 \frac{(1 - e^{-2 \beta \omega})}{(1 + e^{-\beta \omega})^2} (2 - e^{-4 \beta \omega}) Y(t),
\end{align}
with the coefficients (some are given in Eq. (\ref{BB1})) expressed as double integrals over gravitational-wave correlators:  
\begin{equation}\label{eq:coefficients}
\begin{array}{l @{\hspace{1cm}} l}
C(t) = \displaystyle \int_0^t dt_2 \int_0^t dt_1 \langle \hat{\gamma}^I(t_1) \hat{\gamma}^I(t_2) \rangle_G e^{T_-}, \\[12pt]
X(t) = \int_0^t dt_2 \int_0^t dt_1 \langle \hat{\gamma}^I(t_2) \rangle_G \langle \hat{\gamma}^I(t_1) \rangle_G \cos[2 \omega (t_1 + t_2)], \\[12pt]
Y(t) = \displaystyle \int_0^t dt_2 \int_0^t dt_1 \langle \hat{\gamma}^I(t_2) \rangle_G \langle \hat{\gamma}^I(t_1) \rangle_G \cos T, &
\end{array}
\end{equation}
where \(T = 2 \omega (t_1 - t_2)\).

The graviton-induced effects depend explicitly on the two-point correlators \(\langle \hat{\gamma}^I(t_1) \hat{\gamma}^I(t_2) \rangle_G\), which are determined by the initial state of the gravitational field. While a mixed thermal graviton state can, in principle, be represented as a pure state via the thermo-field double construction, it suffices here to evaluate thermal expectation values directly:  
\begin{equation}
    \langle \hat{\gamma}^I(t_2) \hat{\gamma}^I(t_1) \rangle_G = \mathrm{Tr}_{G}\!\left(\hat{\rho}^{g}_{th}(\beta^{'}) \hat{\gamma}^I(t_2) \hat{\gamma}^I(t_1)\right), \quad
    \hat{\rho}^{g}_{th}(\beta^{'}) = \frac{e^{-\beta' \hat{H}_G}}{\mathrm{Tr}_G(e^{-\beta' \hat{H}_G})}.
    \label{g}
\end{equation}  
In realistic experimental or astrophysical scenarios, the detector temperature \(T = 1/\beta\) need not match the graviton bath temperature \(T' = 1/\beta'\). Laboratory detectors are often cryogenically cooled to suppress thermal noise, whereas any ambient graviton population would originate from cosmological or astrophysical sources and is unlikely to be in thermal equilibrium with the detector. Allowing \(T \neq T'\) captures genuinely nonequilibrium dynamics, enables the identification of stimulated graviton processes that vanish in equilibrium, and highlights nonlinear thermal corrections indicative of quantum-gravitational interactions.  
For concreteness, we consider gravitational-wave (GW) modes in a thermal ensemble at temperature \(\beta' = 1/T'\), with the Hamiltonian  
\begin{equation}\label{Hgamma}
    \hat{H}_G = \omega_g \hat{b}_g^\dagger \hat{b}_g,
\end{equation}  
where \(\omega_g\) denotes the GW frequency. Using Eq.~(\ref{h}), the corresponding thermal two-point correlators are given by  
\begin{align}\label{thermal_correlations}
    \langle \hat{\gamma}^I(t_2) \hat{\gamma}^I(t_1) \rangle_G &= C_\gamma^2 \left[ \frac{2 \cos[\omega_g (t_1 - t_2)]}{e^{\beta' \omega_g} - 1} + e^{i \omega_g (t_1 - t_2)} \right], \\
    \langle \hat{\gamma}^I(t_1) \hat{\gamma}^I(t_2) \rangle_G &= C_\gamma^2 \left[ \frac{2 \cos[\omega_g (t_1 - t_2)]}{e^{\beta' \omega_g} - 1} + e^{-i \omega_g (t_1 - t_2)} \right],
\end{align}  
while the one-point function vanishes identically:  
\begin{equation}\label{one_point_zero}
    \langle \hat{\gamma}^I(t_i) \rangle_G = 0.
\end{equation}

\begin{figure}
    \centering
    \includegraphics[width=0.9\linewidth]{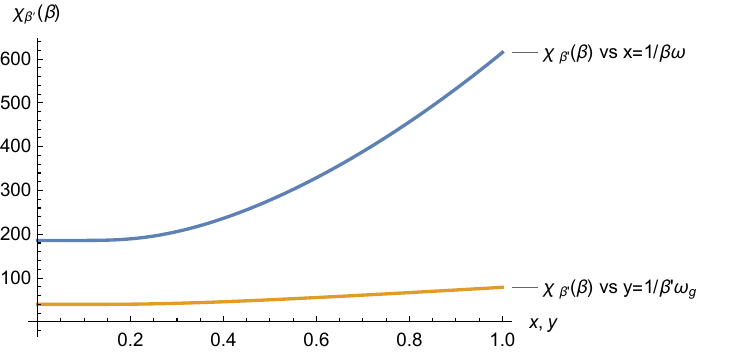}
    \caption{$\chi_{\beta'}(\beta)$ versus both the system and graviton bath temperature depicts purity decreases with rise in temperature suggesting temperature enhances entanglement.}
    \label{TotalPurity}
\end{figure}

Use of the above expressions in (\ref{BB1}) yields the  the explicit forms of the coefficients. 
Here, we give a qualitative plot of $\chi_{\beta'}(\beta)$ versus both $x=\frac{1}{\beta\omega}$ and $y=\frac{1}{\beta'\omega_g}$ (see Eq. (\ref{purity_explicit})) with $\omega=2\pi\times100\mathrm{Hz}$ and $\omega_g=2\pi\times1\mathrm{kHz}$ in Figure    \ref{TotalPurity}. A fixed value $y=5$ is taken when plotting as a function of $x$. Similarly a fixed value $x=20$ is considered when plotting as a function of $y$. For both cases time has been fixed to $t=1.43\mathrm{s}$. The plot clearly shows that both the temperatures of the oscillator-system and graviton bath enhance the entanglement between the thermal vacuum states of the two HOs. Furthermore, mainly the HO bath is influencing the entanglement enhancement, with the graviton bath merely taking a small part in it, as is evident in Figure \ref{TotalPurity}. 

Further substituting the graviton correlators into (\ref{purity_1Q}) yields the explicit purity of the physical oscillator:  
\begin{align}\label{purity_final}
    P^\beta_{1Q}(t) &= \left(\frac{1 - e^{-\beta \omega}}{1 + e^{-\beta \omega}} \right) 
    - \frac{8 C_\gamma^2}{(1 + e^{-\beta \omega})^2 (1 - e^{-\beta \omega})} \Big[ 2(1 + e^{-3 \beta \omega})(1 - e^{-\beta \omega}) + e^{-5 \beta \omega}(2 - e^{-\beta \omega}) \Big] \notag \\
    &\quad \times \left\{ \frac{1}{e^{\beta' \omega_g} - 1} \left[ \frac{\sin^2(\Omega t/2)}{\Omega^2} + \frac{\sin^2(\Omega' t/2)}{\Omega'^2} \right] + \frac{\sin^2(\Omega t/2)}{\Omega^2} \right\} \notag \\
    &\quad + 4 C_\gamma^2 \Bigg[ \frac{4 e^{-2 \beta \omega}}{(1 + e^{-\beta \omega})^2} - \frac{2 e^{-6 \beta \omega}}{(1 + e^{-\beta \omega})^3 (1 - e^{-\beta \omega})} \Big( (2 - e^{-\beta \omega}) - e^{-\beta \omega} (1 - e^{-\beta \omega})^2 \Big) \Bigg] \notag \\
    &\quad \times \left\{ \frac{1}{e^{\beta' \omega_g} - 1} \left[ \frac{\sin^2(\Omega t/2)}{\Omega^2} + \frac{\sin^2(\Omega' t/2)}{\Omega'^2} \right] + \frac{\sin^2(\Omega' t/2)}{\Omega'^2} \right\} \notag \\
    &\quad - 2 C_\gamma^2 \frac{4 e^{-3 \beta \omega}}{(1 + e^{-\beta \omega})^3} (2 - e^{-4 \beta \omega}) \left( \frac{e^{\beta' \omega_g} + 1}{e^{\beta' \omega_g} - 1} \right) \notag \\
    &\quad \times \left[ \frac{\sin^2(\Omega t/2)}{\Omega \Omega'} + \frac{\sin^2(\Omega' t/2)}{\Omega \Omega'} - \frac{\sin^2\left( (\Omega + \Omega') t/2 \right)}{\Omega \Omega'} \right],
\end{align}  
with \(\Omega = 2 \omega + \omega_g\) and \(\Omega' = 2 \omega - \omega_g\).  
Because the one-point function vanishes [Eq.~\eqref{one_point_zero}], the terms \(X(t)\) and \(Y(t)\) drop out, leaving the purity explicitly dependent on both the detector temperature \(\beta\) and graviton temperature \(\beta'\) as well as two point correlation of graviton modes.
Again, we provide a qualitative profile of physical purity varying with both the HO and graviton bath temperatures in Figure \ref{fg} and Figure \ref{gf}. 
\begin{center}
\begin{minipage}{0.8\linewidth}
    \includegraphics[width=\linewidth]{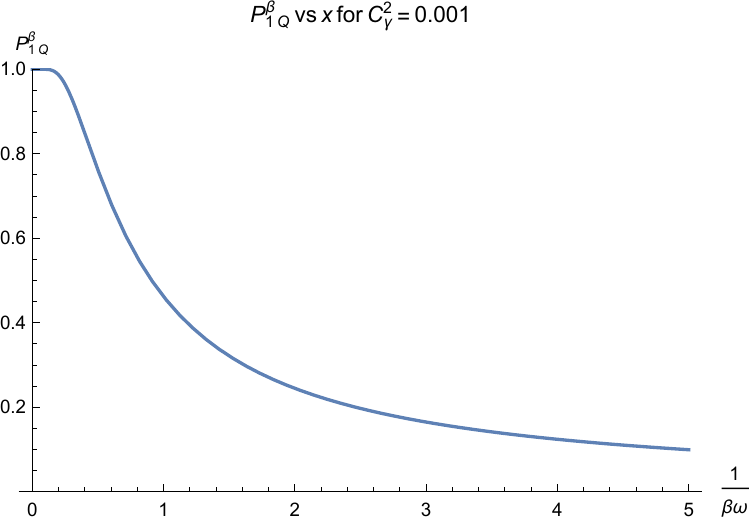}
    \captionof{figure}{$P^\beta_{1Q}(1.43 \mathrm{s})$ vs $x=\frac{1}{\beta\omega}$ plot for $C^2_\gamma=0.001$: Purity decreases with HO bath temperature favoring entanglement.}
    \label{fg}
\end{minipage}
\vskip 5mm
\hfill
\begin{minipage}{0.8\linewidth}
    \includegraphics[width=\linewidth]{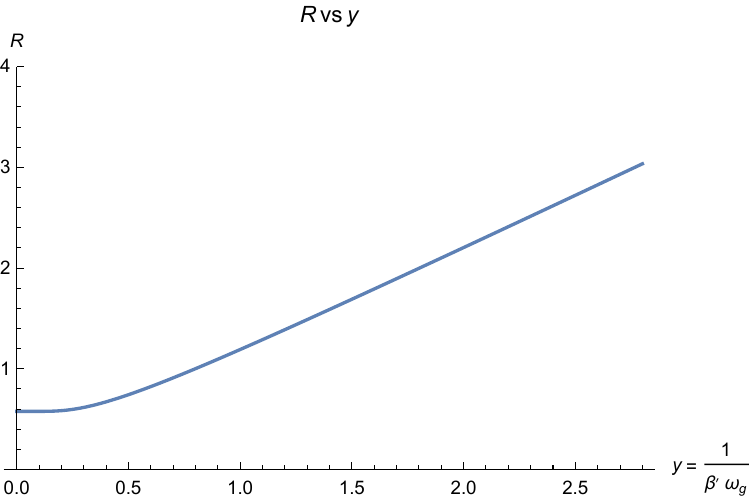}
    \captionof{figure}{$R(y)$ vs $y=\frac{1}{\beta'\omega_g}$ plot: $R(y)$ increases with graviton bath temperature facilitating entanglement enhancement.}
    \label{gf}
\end{minipage}
\end{center}
The values of the parameters $\omega$,~$\omega_g$ and $t$ are chosen to be those considered in Figure \ref{TotalPurity}; i.e. $\omega=2\pi\times100\mathrm{Hz}$, $\omega_g=2\pi\times1\mathrm{kHz}$, and $t=1.43\mathrm{s}$. In Fig. \ref{fg} we fix $y=1/(\beta'\omega_g) = 5$ and plot $P^\beta_{1Q}$ as a function of $x=1/(\beta\omega)$. It shows that the purity decreases with $x$. 
In case of temperature dependence of purity of the physical oscillator due to the graviton bath, we fix the detector-system temperature ($x=20$), keeping $y=\frac{1}{\beta'\omega_g}$ as a variable, thereby 
\begin{equation}
P^\beta_{1Q}(1.43\mathrm{s})=0.025-C^2_\gamma~R(y)\times10^{-9}\mathrm{s}^2~,
\end{equation}
where 
\begin{equation}
R(y)=43\left(\frac{18.20}{e^{\frac{1}{y}}-1}+10.75\right)+13\left(\frac{18.20}{e^{\frac{1}{y}}-1}+7.45\right)-18\left(\frac{e^{\frac{1}{y}}+1}{e^{\frac{1}{y}}-1}\right)~.
\end{equation}
Now we plot $R$ as a function of $y$ in Fig. \ref{gf} which shows that $R$ increases with the increase of $y$. Thus the purity decreases with the increase of graviton temperature.
The nature of plots indicate that both the HO and graviton bath favor entanglement enhancement. 

In the zero-temperature limit of the detector (\(\beta \to \infty\)), the physical purity of the first oscillator simplifies to
\begin{equation}
    P^\infty_{1Q}(t) = 1 - 4 B(t) + 4 Y(t) = 1 - 2 K^{(2)} + 2 K''^{(2)},
\end{equation}
where
\begin{equation}
    K''^{(2)} = 2 \int_0^t dt_2 \int_0^t dt_1 \, \langle \hat{\gamma}^I(t_2) \rangle_G \langle \hat{\gamma}^I(t_1) \rangle_G \, \cos T.
\end{equation}
As shown in \cite{Nandi:2024jyf}, $K''^{(2)}$ vanishes when the expectation value of the gravitational-wave operator is zero, $\langle \hat{\gamma}^I(t) \rangle_G = 0$, corresponding physically to the graviton vacuum state. 
The same holds in a thermal state, since it is diagonal in the number basis and invariant under parity; because $\hat{\gamma}^I(t)$ is parity-odd, its expectation value must vanish, leaving only correlation functions as physically relevant quantities.

For the intermediate purity $P_Q^\beta(t)$, which retains the first detector's fictitious-mode degrees of freedom, the zero-temperature limit of the detector reduces to
\begin{align}\label{PQinfty}
    P_Q^\infty(t) &= 1 - 8 \int_0^t dt_2 \int_0^t dt_1 \, \langle \hat{\gamma}^I(t_2) \hat{\gamma}^I(t_1) \rangle_G e^{T_-} + 4 \int_0^t dt_2 \int_0^t dt_1 \, \langle \hat{\gamma}^I(t_2) \rangle_G \langle \hat{\gamma}^I(t_1) \rangle_G e^{T_-} \notag \\
    &= 1 - 4 K^{(2)} + 4 K'^{(2)},
\end{align}  
where  
\begin{equation}
K^{(2)} = 2 \int_0^t dt_2 \int_0^t dt_1 \, \langle \hat{\gamma}^I(t_2) \hat{\gamma}^I(t_1) \rangle_G e^{T_-}, \quad
K'^{(2)} = \int_0^t dt_2 \int_0^t dt_1 \, \langle \hat{\gamma}^I(t_2) \rangle_G \langle \hat{\gamma}^I(t_1) \rangle_G e^{T_-},
\end{equation}
with \(T_- = 2 i \omega (t_1 - t_2)\). In \cite{Nandi:2024jyf}, \(K'^{(2)}\) vanishes under the assumption that the graviton is in its pure vacuum state. Moreover, as we have noticed, this also vanishes for thermal state of graviton (\ref{g}) as well since $\langle \hat{\gamma}^I(t_i) \rangle_G = 0$ even in the number basis of graviton state $\{\ket{n_G}\}$. Then, in presence of the graviton thermal bath we get 
\begin{equation}
    K^{(2)} =8C_\gamma^2\Big[\frac{1}{e^{\beta'\omega_g}-1}\Big\{\frac{\sin^2(\Omega t/2)}{\Omega^2}+\frac{\sin^2(\Omega' t/2)}{\Omega'^2}\Big\}+\frac{\sin^2(\Omega t/2)}{\Omega^2}\Big],\qquad K'^{(2)} =0
\end{equation}
thereby reducing Eq. \eqref{PQinfty} to $P_Q^\infty(t) =1 - 4 K^{(2)}$~.
Finally, in the double zero-temperature limit (\(\beta, \beta' \to \infty\)), the physical purity further reduces to
\begin{equation}
    P^\infty_{1Q}(t) = 1 - 2 K^{(2)}(t) = 1 - 16 C_\gamma^2 \frac{\sin^2(\Omega t/2)}{\Omega^2} ,
\end{equation}  
recovering the vacuum result reported in \cite{Nandi:2024jyf}.

\subsection{Entanglement Entropy}
In a multimode quantum system, gravitational waves can induce entanglement between different subsystems. A standard way to quantify this is through the \textit{entanglement entropy} of one subsystem’s reduced state.  
Here, we focus on the entanglement between the first thermal harmonic oscillator (HO) and the remaining unobserved modes, taking into account both the thermal state of the system and the influence of gravitational waves.
To characterize this entanglement, we use the \textit{quantum Rényi entropy}, a generalization of the von Neumann entropy~\cite{Renyi1970}, defined as  
\begin{equation}\label{eq:renyi_entropy}
    S_{\alpha}(\hat{\rho}) = \frac{1}{1-\alpha} \ln{\mathrm{Tr}(\hat{\rho}^\alpha)}, \quad \alpha \in (0,1) \cup (1,\infty)~,
\end{equation}
for a given state $\hat{\rho}$.
In the limit $\alpha \to 1$, $S_\alpha$ reduces to the von Neumann entropy.  
For the special case $\alpha = 2$, we obtain the \textit{second-order Rényi entropy}:
\begin{equation}\label{eq:renyi_second_order}
    S_2(\hat{\rho}) = -\ln{\mathrm{Tr}(\hat{\rho}^2)}~.
\end{equation}

The reduced density matrix $\hat{\rho}^\beta_{1\tilde{1}}(t)$  of the first physical oscillator and its TFD partner is given by (\ref{thermal_density_matrix_final_compact})
and the corresponding second-order Rényi entropy is
\begin{equation}\label{eq:S2_1t1_def}
    S^{\beta}_{2(1\tilde{1})}(t) 
    \;=\; -\ln \mathrm{Tr}_{1,\tilde{1}}\!\left[\big(\hat{\rho}^\beta_{1\tilde{1}}(t)\big)^2\right]
    \;=\; -\ln P^\beta_{Q}(t),
\end{equation}
On the other hand we can also evaluate the second-order Rényi entropy for the reduced state (\ref{reduced_density_matrix}) of the \textit{first harmonic oscillator} in the presence of both thermal effects and gravitational wave interactions.  
Keeping terms only up to order $\mathcal{O}(\gamma^2)$, the entropy is given by
\begin{equation}\label{eq:entropy_physical_system}
    S^\beta_{2(1)}(t) = -\ln{\mathrm{Tr}_1 \left[(\hat{\rho}_1^\beta(t))^2\right]} = -\ln{ P^\beta_{1Q}(t)}~,
\end{equation}
where the first index ``2" indicates $\alpha = 2$ (second-order Rényi entropy), the second index ``1" denotes the reduced subsystem corresponding to the first HO. 
Note that the entanglement entropies are directly connected to respective purities. Therefore the conclusions from the nature of purities, which we discussed in last subsection, are also trivially implied through the entropies. Hence we refrain from repeating those again.




Having quantified the entanglement induced by gravitational waves via the Rényi entropy of the first oscillator, we now shift focus to a related but complementary observable: the local thermal response of the oscillator mode to a quantum gravitational wave bath. This perspective reveals not only global entanglement properties but also the dynamical occupation number behavior and quantum memory effects arising from detector–graviton interactions.

\section{Quantum gravitational wave–induced thermal response}\label{sec:therm-respons}
At this stage it is important to emphasize that the background temperature alone cannot generate entanglement between the two independent harmonic oscillator (HO) detector modes. The entanglement dynamics are entirely driven by the interaction with gravitational waves (GWs). This setting therefore provides a natural framework to explore both the global entanglement evolution and the local thermal response of a single HO mode to quantum GWs.

To quantify this local response we consider the expectation value of the number operator for the first mode,
\begin{equation}
\langle \hat{N}_1 \rangle_{\beta,\beta'} = \mathrm{Tr}_1 \!\left( \hat{N}_1 \hat{\rho}^\beta_1(t) \right),
\quad \hat{N}_1 = \hat{a}^\dagger_1 \hat{a}_1,
\end{equation}
where $\hat{\rho}^\beta_1(t)$ is the reduced density matrix of the first HO mode (see Eq.~\eqref{reduced_density_matrix}), initially at inverse temperature $\beta$, while $\beta'$ denotes the inverse temperature of the graviton bath. Evaluating this trace in the HO eigenbasis and substituting the explicit coefficients from Eqs.~\eqref{reduced_density_matrix} as well as using \eqref{eq:coefficients}, we obtain

\begin{equation}\label{distribution_final_compact}
\begin{split}
\langle \hat{N}_1 \rangle_{\beta,\beta'} 
&= \langle \hat{n}_\beta \rangle + \langle \hat{n}_\gamma \rangle \, G_1(\langle \hat{n}_\beta \rangle) + \langle \hat{n}_{\beta'} \rangle \langle \hat{n}_\gamma \rangle \, G_2(\langle \hat{n}_\beta \rangle) \\
&\quad + 2K_1^{(2)}(t) \Big[ \langle \hat{n}_{\beta'} \rangle \, G_3(\langle \hat{n}_\beta \rangle) - G_4(\langle \hat{n}_\beta \rangle) \Big] \\
&\quad + 2K_2^{(2)}(t) \Big[ G_5(\langle \hat{n}_\beta \rangle) + \langle \hat{n}_{\beta'} \rangle \, G_6(\langle \hat{n}_\beta \rangle) \Big].
\end{split}
\end{equation}
The polynomials $G_i$, appeared above, are defined as
\begin{align}
G_1(\langle \hat{n}_\beta \rangle) &= 1 + \left(2 - \frac{\Omega}{\Omega'}\right) \langle \hat{n}_\beta \rangle + \left(1 - \frac{5\Omega}{2\Omega'}\right) \langle \hat{n}_\beta \rangle^2 - \frac{3\Omega}{2\Omega'} \langle \hat{n}_\beta \rangle^3, \\
G_2(\langle \hat{n}_\beta \rangle) &= 1 + 2\left(1 - \frac{\Omega}{\Omega'}\right) \langle \hat{n}_\beta \rangle - \frac{5\Omega}{\Omega'} \langle \hat{n}_\beta \rangle^2 - \frac{3\Omega}{\Omega'} \langle \hat{n}_\beta \rangle^3, \\
G_3(\langle \hat{n}_\beta \rangle) &= 1 + 2\left(1 - \frac{\Omega'}{\Omega}\right) \langle \hat{n}_\beta \rangle - \frac{5\Omega'}{\Omega} \langle \hat{n}_\beta \rangle^2 - \frac{3\Omega'}{\Omega} \langle \hat{n}_\beta \rangle^3, \\
G_4(\langle \hat{n}_\beta \rangle) &= \frac{\Omega'}{\Omega} \langle \hat{n}_\beta \rangle + \left(1 + \frac{5\Omega'}{2\Omega}\right) \langle \hat{n}_\beta \rangle^2 + \frac{3\Omega'}{2\Omega} \langle \hat{n}_\beta \rangle^3, \\
G_5(\langle \hat{n}_\beta \rangle) &= \langle \hat{n}_\beta \rangle + \frac{5}{2} \langle \hat{n}_\beta \rangle^2 + \frac{3}{2} \langle \hat{n}_\beta \rangle^3, \\
G_6(\langle \hat{n}_\beta \rangle) &= 2 \langle \hat{n}_\beta \rangle + 5 \langle \hat{n}_\beta \rangle^2 + 3 \langle \hat{n}_\beta \rangle^3,
\end{align}
 The quantities 
\begin{equation}
\langle \hat{n}_\beta\rangle = \frac{1}{e^{\beta\omega} - 1}, 
\quad \langle \hat{n}_{\beta'}\rangle = \frac{1}{e^{\beta'\omega_g} - 1},
\quad \langle \hat{n}_\gamma\rangle = \frac{16C_\gamma^2 \sin^2(\Omega t/2)}{\Omega^2},
\end{equation}
denote, respectively, the detector thermal occupation, the thermal graviton occupation, and the GW-induced excitation factor, while
\begin{equation}
K_1^{(2)}(t) = \frac{8C_\gamma^2 \sin^2(\Omega' t/2)}{\Omega'^2}, 
\quad K_2^{(2)}(t) = \frac{8C_\gamma^2 \sin^2(2\omega t)}{\Omega\Omega'},
\quad \Omega = 2\omega + \omega_g, \quad \Omega' = 2\omega - \omega_g,
\end{equation}
encode phase-dependent correlations.

\subsection{Structure of the quantum GW contribution}

The structure of Eq.~\eqref{distribution_final_compact} reveals that the detector excitation is not a simple thermal distribution but rather a superposition of qualitatively distinct contributions:

\begin{enumerate}
\item \textbf{Baseline thermal response:}  
The first term, $\langle \hat{n}_\beta \rangle$, corresponds to the equilibrium Bose–Einstein distribution of the oscillator in the absence of any GW coupling. It sets the thermal baseline against which all GW-induced effects must be compared.

\item \textbf{Vacuum graviton fluctuations:}  
The second term, $\langle \hat{n}_\gamma\rangle\, G_1(\langle \hat{n}_\beta\rangle)$, survives even in the zero-temperature limit of the graviton bath ($\beta'\to\infty$). Its origin lies in the vacuum fluctuations of the quantized GW field, which induce spontaneous detector excitations. This term therefore represents a genuine quantum signature with no classical analogue.

\item \textbf{Thermal graviton enhancement:}  
The third term, $\langle \hat{n}_{\beta'} \rangle \langle \hat{n}_\gamma \rangle\, G_2(\langle \hat{n}_\beta\rangle)$, vanishes when the graviton bath is in vacuum but becomes relevant for finite $\beta'$. Its proportionality to $\langle \hat{n}_{\beta'}\rangle$ identifies it with stimulated absorption and emission processes driven by a thermal graviton population, in direct analogy with detailed balance in standard quantum optics.

\item \textbf{Phase-coherent correlations:}  
Finally, the terms proportional to $K_1^{(2)}(t)$ and $K_2^{(2)}(t)$ oscillate at shifted frequencies $\Omega'$ and $2\omega$, respectively. Their explicit time dependence encodes phase-sensitive correlations between the detector and graviton modes. These contributions modulate both vacuum and thermal graviton effects, leading to interference-like features and memory effects in the detector response.
\end{enumerate}
Taken together, this classification shows that the detector’s thermal response is composed of four distinct layers: a purely thermal baseline, a quantum vacuum contribution, thermal graviton enhancement, and coherent dynamical modulations. This rich structure carries unambiguous imprints of the quantized nature of the GW field. 
We next explore the dynamical consequences of these quantum gravitational wave contributions, focusing on the emergence of a \emph{quantum memory effect} and a time-crystal-like phase in the detector’s dynamics.

\subsection{Quantum-memory and time-crystal-like behavior induced by gravitons}

In contrast to the classical scenario, where finite-temperature or classically polarized gravitational waves (GWs) cannot sustain entanglement between detector modes, quantized gravitational waves generate qualitatively new dynamical effects. A purely classical GW may transiently modulate detector occupations, but the subsystem rapidly relaxes to equilibrium once the external drive is oscillates to zero (as the gravitational wave modes are taken as sinusoidal), producing only short-lived memory. By contrast, in the quantum case, persistent correlations survive due to vacuum fluctuations and entanglement with graviton modes, marking a genuinely quantum phenomenon.

To make this precise, we quantify deviations from equilibrium as
\begin{equation}
\delta \hat{\rho}^{\beta,\beta'}_g(t) = \hat{\rho}_1^\beta(t) - \hat{\rho}^\beta_{1,\mathrm{eq}},
\end{equation}
where $\hat{\rho}^\beta_{1,\mathrm{eq}}$ is the stationary thermal state at inverse temperature $\beta$. The instantaneous deviation in occupation number is
\begin{equation}
\Delta {N}_1(t) \equiv \langle \hat{N}_1 \rangle_{\beta,\beta'} - \langle \hat{n}_\beta \rangle,
\end{equation}
which encodes the nonequilibrium response of the detector. From Eq.~\eqref{distribution_final_compact}, oscillatory terms are driven by graviton vacuum fluctuations, involving $\langle \hat{n}_\gamma \rangle$, $K_1^{(2)}(t)$, and $K_2^{(2)}(t)$. Remarkably, these oscillations persist even for a pure graviton vacuum ($\langle \hat{n}_{\beta'} \rangle=0$), reflecting entanglement between detector and graviton modes. This defines the {\it graviton-induced quantum memory effect}:
\begin{equation}\label{distribution_final_compact1}
\begin{split}
\Delta {N}_1(t) 
&=  \langle \hat{n}_\gamma \rangle \, G_1(\langle \hat{n}_\beta \rangle) + \langle \hat{n}_{\beta'} \rangle \langle \hat{n}_\gamma \rangle \, G_2(\langle \hat{n}_\beta \rangle) \\
&\quad + 2K_1^{(2)}(t) \Big[ \langle \hat{n}_{\beta'} \rangle \, G_3(\langle \hat{n}_\beta \rangle) - G_4(\langle \hat{n}_\beta \rangle) \Big] \\
&\quad + 2K_2^{(2)}(t) \Big[ G_5(\langle \hat{n}_\beta \rangle) + \langle \hat{n}_{\beta'} \rangle \, G_6(\langle \hat{n}_\beta \rangle) \Big].
\end{split}
\end{equation}
which has no analogue in a purely classical GW background.

The oscillations of $\langle \hat{N}_1(t) \rangle$ are built from three distinct frequencies:
\begin{equation}
\Omega = 2\omega + \omega_g, \quad \Omega' = 2\omega - \omega_g, \quad 2\omega,
\end{equation}
corresponding to the detector's internal response. The appearance as well as interpretation of these frequencies in different terms of (\ref{distribution_final_compact1}) is being explicitly mentioned in Table \ref{tab:frequencies}. For the signal to repeat after some time $\tau$, all three oscillatory factors must simultaneously align:
\begin{equation}
\Omega \tau = 2\pi k, \quad \Omega' \tau = 2\pi \ell, \quad (2\omega) \tau = 2\pi m,
\end{equation}
for integers $k, \ell, m$. Defining $r = \omega_g / \omega$, these frequencies become
\begin{equation}
\Omega = (2+r)\omega, \quad \Omega' = (2-r)\omega, \quad 2\omega,
\end{equation}
and the alignment condition reduces to requiring all ratios among $\Omega$, $\Omega'$, and $2\omega$ to be rational:
\begin{equation}
\frac{\Omega}{\Omega'} = \frac{2+r}{2-r}, \quad \frac{\Omega}{2\omega} = \frac{2+r}{2}, \quad \frac{\Omega'}{2\omega} = \frac{2-r}{2}.
\end{equation}
Thus, if $r$ is rational, the detector dynamics lock into a strictly periodic orbit with period determined by the least common multiple of the three oscillatory timescales. If $r$ is irrational, quasi-periodic behavior arises. In both cases, the subsystem resists thermalization: memory effects due to entanglement with the graviton field maintain ordered oscillations for times much longer than the GW period $T_g = 2\pi/\omega_g$.

\begin{table}[h!]
\centering
\begin{tabular}{|c|p{6cm}|p{5cm}|}
\hline
\textbf{Frequency} & \textbf{Origin / Physical Interpretation} & \textbf{Contribution in $\langle N_1(t) \rangle$} \\
\hline
$\Omega = 2\omega + \omega_g$ & Detector’s internal response enhanced by GW absorption & Appears in $\langle \hat{n}_\gamma \rangle G_1$ \\
\hline
$\Omega' = 2\omega - \omega_g$ & Detector’s internal response with GW emission & Appears in $K_1^{(2)}(t) G_4$ \\
\hline
$2\omega$ & Detector’s intrinsic oscillation frequency (without GW) & Appears in $K_2^{(2)}(t) G_5$ \\
\hline
\end{tabular}
\caption{Relevant oscillation frequencies of the detector and their contributions to the graviton-induced quantum memory.}
\label{tab:frequencies}
\end{table}

As a consequence of the interaction with the graviton field, the detector’s reduced dynamics obey
\begin{equation}
\Delta N_1 (t+T_g) \neq\Delta N_1 (t),
\quad T_g = \tfrac{2\pi}{\omega_g},
\end{equation}
even for a perfectly monochromatic GW. This inequality highlights that, while the full system (detector plus graviton field) continues to respect the Hamiltonian’s periodicity, the reduced mixed state $\hat{\rho}^\beta_1(t)$ evolves at shifted frequencies. The apparent breaking of time-translation symmetry is therefore not fundamental but an effective phenomenon, arising from subsystem–bath entanglement and the associated quantum memory. It manifests as oscillatory relaxation toward equilibrium, characterized by collapses and revivals rather than simple monotonic damping, with amplitude and period set by the quantum contributions $\langle \hat{n}_\gamma \rangle$ and $K_{1,2}^{(2)}(t)$.

In fact, higher detector temperatures influence the persistence of these  oscillations but do not independently generate memory. Over long timescales, the subsystem gradually approaches thermal equilibrium as correlations with the graviton field spread throughout the bath. However, the transient \textit{prethermal time-crystal}-like phase---with its characteristic revivals---persists over many periods of the GW drive before full thermalization is achieved. This description clarifies that the ``relaxation'' is inherently oscillatory, reflecting the coherent nature of graviton-induced correlations rather than irreversible damping.

In summary, quantized gravitational wave modes---whether in vacuum or thermal states---induce a new form of GW memory absent in classical theory. This graviton-induced quantum memory manifests as a prethermal time-crystal-like phase in detector dynamics, arises from subsystem--bath entanglement, and constitutes a clear signature of the quantum nature of gravitons.


\

\section{Conclusions}\label{sec:conclu}

In this work, we develop a framework to study how the plus-polarized mode of 
\emph{quantized} gravitational waves (GWs) interacts with finite-temperature detectors 
modeled as two independent harmonic oscillators, with no direct coupling between the 
detector modes. The detector dynamics are formulated within the thermo-field dynamics 
(TFD) approach, which enables a unitary operatorial treatment of thermal averages, 
while the graviton sector is described using thermal ensemble correlators at 
temperature \(T'\). The perturbative time evolution is obtained via the Magnus 
expansion, and reduced density matrices are derived by tracing over unobserved 
degrees of freedom. This formulation is particularly advantageous as it separates the 
effects of the detector temperature \(T\) from those of the graviton bath \(T'\), 
allowing us to evaluate purity, R\'enyi entropy, and occupation numbers, and to 
distinguish global entanglement from local thermal response.  

Our analysis leads to two central findings:  

\begin{itemize}
    \item \textbf{Graviton-induced quantum memory} – Detector excitations retain a persistent offset even after the GW drive has ceased. Unlike in the classical case, where memory effects could arise only as a thermal modulation of externally imposed oscillations and decay once the drive vanishes, here the effect persists even in the zero-temperature graviton vacuum. This persistence is a direct manifestation of graviton quantization and provides a testable signature of quantum gravity.  

    \item \textbf{Prethermal time-crystal--like phase} – The subsystem develops oscillations at internal frequencies  
    \begin{equation}
        \Omega = 2\omega \pm \omega_g ,
    \end{equation}
    distinct from the GW drive frequency. This leads to reduced-level time-translation symmetry breaking that cannot be removed by any unitary transformation within the detector Hilbert space. Importantly, while classical GWs merely impose periodicity from the outside, quantized gravitons generate this time-crystal--like behavior intrinsically, through vacuum fluctuations and entanglement with the detector modes.  
\end{itemize}

These results highlight a sharp contrast with our earlier analysis \cite{Dutta:2025bge} of classical gravitational waves with cross polarization, where entanglement between the two detector oscillator modes arose due to their direct interaction mediated by the classical GW modes. In that case, the detector temperature \(\beta\) acted only as a catalyst that modified thermal distributions (mixing Bose--Einstein and Maxwell--Boltzmann statistics) but could not generate entanglement on its own; rather, entanglement was harvested from the classical GWs. By contrast, the classical plus-polarized mode of gravitational waves does not generate any entanglement. In the quantized case, however, entanglement and memory are unavoidable, and the graviton bath associated with the plus-polarized modes, at temperature \(\beta'\), introduces nonlinear fingerprints---up to cubic corrections in occupation number---that further magnify both memory and temporal order. This separation of roles between \(\beta\) and \(\beta'\) provides a unique diagnostic for distinguishing quantum from classical gravitational imprints.  

From the perspective of detectability, our analysis suggests concrete experimental possibilities. Because the memory effect survives even in the graviton vacuum, it provides a testable signature of quantization that cannot be explained by classical GWs. Optomechanical and interferometric platforms already operating near the quantum regime could probe these signatures through persistent offsets in detector occupation and oscillations at shifted internal frequencies. Tabletop resonators, cryogenically cooled oscillators, and future space-based missions such as LISA offer promising testbeds. In addition, nonclassical graviton states may be probed using higher-order correlation techniques, such as Hanbury--Brown--Twiss interferometry~\cite{HanburyBrown:1956bqd,Brown:1956zza} with next-generation technology~\cite{PhysRevX.15.011034,61j9-cjkk}, potentially revealing squeezed or squeezed-thermal primordial GWs from the early universe. 

These theoretical signatures align naturally with recent proposals for probing the quantum structure of gravitational radiation. In particular, Manikandan and Wilczek~\cite{Manikandan2025, Manikandan:2025hlz} have outlined concrete operational tests—embodied in the \emph{coherent-state hypothesis}—that employ counting statistics in resonant bar detectors and phase-sensitive (homodyne or heterodyne) measurements to distinguish coherent (classical-like) gravitational radiation from thermal or squeezed quantum states. The graviton-induced memory offsets and subharmonic oscillations identified here correspond directly to deviations that would appear as excess noise or sideband structures in those counting and quadrature observables. Our analysis therefore complements these experimental strategies by specifying the microscopic origin and parametric dependence of such deviations, linking measurable detector responses to the underlying quantum state of the gravitational field.

In summary, we have shown that quantized GWs leave behind unambiguous, state-dependent quantum signatures—persistent graviton-induced memory, a prethermal time-crystal–like phase, and nonlinear thermal amplification—that cannot be mimicked by classical gravitational waves. These effects not only sharpen the theoretical distinction between classical and quantum gravity but, together with emerging experimental probes such as those proposed in~\cite{Manikandan2025, Manikandan:2025hlz}, open realistic avenues for tabletop-scale tests of the quantum nature of gravitons.

\acknowledgments

MD gratefully acknowledges funding support from the Indian Institute of Technology Guwahati, India, through the Junior Research Fellowship (JRF) program. PN appreciates the support provided by the National Institute for Theoretical and Computational Sciences (NITheCS) via the Rector’s Postdoctoral Fellowship Program (RPFP). He also extends sincere thanks to Prof. Frederik G. Scholtz, Prof. Igor Pikovski, and Prof. Kazuhiro Yamamoto, as well as NITheCS Director Prof. Francesco Petruccione, for their valuable discussions and continuous academic guidance. BRM expresses gratitude to the people of India for their unwavering support of fundamental scientific research.

\begin{appendix}
\section*{Appendices}

\appendix

\section{Derivation of the Hamiltonian for graviton modes}
\label{Appendix A}
The analysis is being followed from \cite{Parikh:2020fhy,Nandi:2024jyf}
We start from linearized gravity about Minkowski space,
\begin{equation}
g_{\mu\nu}=\eta_{\mu\nu}+h_{\mu\nu}, \qquad |h_{\mu\nu}|\ll 1,
\end{equation}
and impose the transverse--traceless (TT) gauge on the spatial components:
\begin{equation}
h_{0\mu}=0,\qquad \partial^i h_{ij}=0,\qquad h^i{}_{i}=0.
\end{equation}
In this gauge the quadratic (free) gravitational (Einstein-Hilbert) action is
\begin{equation}
S_{\mathrm{grav}}=-\frac{1}{64\pi G}\int d^4x\,(\partial_\alpha h_{ij})(\partial^\alpha h^{ij})
=\frac{1}{64\pi G}\int d^4x\,\Big(\dot h_{ij}\dot h^{ij}-\partial_k h_{ij}\,\partial_k h^{ij}\Big).
\end{equation}

To exhibit the canonical variables, we decompose the metric deformation into TT plane waves in a cubic box of side $L$ (volume $V=L^3$):
\begin{equation}
h_{ij}(t,\vec x)=\frac{1}{\sqrt{\hbar G}}\sum_{\vec k,\,s=\{+,\times\}} 
q_{\vec k,s}(t)\,e^{i\vec k\cdot\vec x}\,\epsilon^{(s)}_{ij}(\vec k),
\end{equation}
where $\epsilon^{(s)}_{ij}(\vec k)$ are real, transverse and traceless polarization tensors, orthonormal on the TT subspace. 
The ``canonical coordinate" for each graviton degree of freedom is the TT-projected Fourier amplitude of the metric deformation:
\begin{equation}
q_{\vec k,s}(t)=\sqrt{\hbar G}\,\frac{1}{V}\int d^3x\;e^{-i\vec k\cdot\vec x}\,
\epsilon^{(s)}_{ij}(\vec k)\,h_{ij}(t,\vec x),
\end{equation}
so that the metric perturbation is reconstructed from the set $\{q_{\vec k,s}\}$ by the inverse expansion above. 
Reality of $h_{ij}$ implies $q^{\ast}_{\vec k,s}(t)\epsilon^{(s)}_{ij}(\vec k)=q_{-\vec k,s}(t)\epsilon^{(s)}_{ij}(-\vec k)$. 

Substituting the mode expansion and using 
$\int_V d^3x\,e^{i(\vec k-\vec k')\cdot\vec x}=V\,\delta_{\vec k,\vec k'}$ 
and polarization orthonormality, the action becomes a sum of decoupled oscillators: 
\begin{equation}
S = \frac{L^3}{32\pi\,\hbar\,G^2} \int dt \sum_{\vec k, s} \Big( \dot q_{\vec k,s}^2 - \omega_k^2 \, q_{\vec k,s}^2 \Big) \equiv \int dt~ \mathcal{L}, 
\qquad \omega_k \equiv |\vec k|.
\end{equation}
Here, the characteristic length scale \(L\) corresponds to the gravitational wavelength (with $c=1$):
\begin{equation}
L = \frac{1}{\omega_g}.
\end{equation}
The canonical momentum conjugate to \(q_{\vec k,s}\) is
\begin{equation}
p_{\vec k,s} \equiv \frac{\partial \mathcal{L}}{\partial \dot q_{\vec k,s}} = \frac{L^3}{16\pi\,\hbar\,G^2} \, \dot q_{\vec k,s},
\end{equation}
so that
\begin{equation}
\dot q_{\vec k,s} = \frac{16 \pi \, \hbar \, G^2}{L^3} \, p_{\vec k,s}, 
\quad \text{or equivalently} \quad 
p_{\vec k,s} = \frac{L^3}{16 \pi \, \hbar \, G^2} \frac{d}{dt} q_{\vec k,s}.
\end{equation}
It is now convenient to introduce a mode-independent ``mass'' parameter
\begin{equation}
m \equiv \frac{L^3}{16 \pi \, \hbar \, G^2},
\end{equation}
so that the Hamiltonian for each mode takes the standard harmonic-oscillator form:
\begin{equation}
H_{\vec k,s} = \frac{p_{\vec k,s}^2}{2 m} + \frac{1}{2} m \, \omega_k^2 \, q_{\vec k,s}^2, 
\qquad H = \sum_{\vec k, s} H_{\vec k,s}.
\end{equation}
For concreteness, let us focus on a single mode with wave vector along the $+\hat{z}$ direction and frequency
\begin{equation}
\omega_k := \omega_g = |\vec k|,
\end{equation}
restricted to the $+$ polarization for simplicity (the $\times$ polarization being dynamically identical at the free-field level).
In this case, the effective action and Hamiltonian reduce to
\begin{equation}
S_{\omega_g} = \int dt \, \frac{1}{2} m \left( \dot q_{+}^{2} - \omega_g^2 q_{+}^{2} \right),
\qquad
H_{\omega_g} = \frac{p_{+}^2}{2m} + \frac{1}{2} m \omega_g^2 q_{+}^{2},
\end{equation}
where, for brevity, we have denoted $q_{\vec{k},+} \to q_{+}$ and $p_{\vec{k},+} \to p_{+}$.
Canonical quantization promotes $(q_{+},p_{+})$ to operators with $[\hat q_{+},\hat p_{+}] = i\hbar$. Introducing the ladder operators
\begin{equation}
\hat b_g = \sqrt{\frac{m\omega_g}{2\hbar}}~ \hat q_{+} + \frac{i}{\sqrt{2m\hbar\omega_g}}~ \hat p_{+},
\qquad
\hat b_g^\dagger = \sqrt{\frac{m\omega_g}{2\hbar}}~ \hat q_{+} - \frac{i}{\sqrt{2m\hbar\omega_g}}~ \hat p_{+},
\end{equation}
which satisfy $[\hat b_g, \hat b_g^\dagger] = 1$, the Hamiltonian becomes
\begin{equation}
\hat H_G \equiv \hat H_{\omega_g} = \hbar \omega_g \left( \hat b_g^\dagger \hat b_g + \tfrac{1}{2} \right).
\end{equation}
Thus, each TT graviton mode $(\vec k, s)$ corresponds to a quantum harmonic oscillator of frequency $\omega_g = |\vec k|$, with canonical coordinate given by the TT-projected Fourier amplitude of the metric deformation and canonical momentum proportional to its time derivative, as defined above.

As we have chosen only the plus polarization for the GW, for each mode of the GW i.e. for each $\vec{k}$, we can have
\begin{equation}\label{mode}
    h_{ij}(t)=2\chi(t)\epsilon_+\sigma_{ij}^3=\frac{1}{\sqrt{\hbar G}}q_{+}(t)e^{i\vec{k}.\vec{x}}\epsilon_{ij}^{+}(\vec{k})~,
\end{equation}
where  $\epsilon_{ij}^{+}(\vec{k})$ is the polarization tensor corresponding to the plus polarization. Then from \eqref{mode}, we get
\begin{equation}
    \chi(t)=\frac{1}{2\sqrt{\hbar G}}q_{+}(t)e^{i\vec{k}.\vec{x}}~,
\end{equation}
which for GWs moving along z-direction reduces to $\chi(t)=\frac{1}{2\sqrt{\hbar G}}q_{+}(t)e^{ikz}$. Again, we have already promoted $q_+\rightarrow\hat{q}_+$ i.e to its quantum status. Then, we can expand $\hat{q}_+$ in terms of its modes as: $
\hat{q}_+=\sqrt{\frac{\hbar}{2m\omega_g}}(\hat{b}_ge^{-i\omega_gt}+\hat{b}_g^\dagger e^{i\omega_gt})$. Then, elevating $\chi(t)$ to its corresponding operator and without loss of generality choosing $z=0$, we get
\begin{equation}
    \hat{\chi}(t)=\frac{1}{2\sqrt{\hbar G}}\sqrt{\frac{\hbar}{2m\omega_g}}(\hat{b}_ge^{-i\omega_gt}+\hat{b}_g^\dagger e^{i\omega_gt})~.
\end{equation}
This will yield \eqref{h}.

\section{Evaluation of Eq. \eqref{eq:thermal_state_compact}}\label{Appendix C}
The interaction between the graviton bath and the detector system, being weak,  is best studied in the interaction picture.  Here, the time-evolution operator $\hat{U}^\gamma_{\mathrm{int}}(t)$ given by \eqref{U_Magnus} time evolves the initial state of the detector-graviton system i.e, $ \ket{0,0;G}_{\beta(t=0)}$ perturbatively upto second order in $\gamma$ through \eqref{t-evolved-state}.
Using this the time evolved thermal ground state in interaction picture is given as (terms upto $\mathcal{O}(\gamma^2)$ are kept)
\begin{eqnarray}\label{Int picture g.s}
&&\ket{0,0;G;t}_{\beta}^I=(1-e^{-\beta\omega})\sum_{n_x,n_y}e^{-n_x\beta\omega/2}e^{-n_y\beta\omega/2}\Bigg[\ket{n_x,n_y,\Tilde{n}_x,\Tilde{n}_y}+\alpha^{(1)}_1\ket{n_x+2,n_y,\Tilde{n}_x,\Tilde{n}_y}
\nonumber
\\
&+&\alpha^{(1)}_2\ket{n_x,n_y,\Tilde{n}_x+2,\Tilde{n}_y}+\alpha^{(1)}_3\ket{n_x,n_y+2,\Tilde{n}_x,\Tilde{n}_y}+\alpha^{(1)}_4\ket{n_x,n_y,\Tilde{n}_x,\Tilde{n}_y+2}
\nonumber
\\
&+&\int_0^t dt_1\int_0^{t_1} dt_2\Big(8i\gamma^I(t_1)\gamma^I(t_2)(n_x+n_y+1)\sin{T}\ket{n_x,n_y,\Tilde{n}_x,\Tilde{n}_y}
\nonumber
\\
&+&\frac{1}{2}\comm{\gamma^I(t_1)}{\gamma^I(t_2)}\Big\{ d_1\left(\ket{n_x+4,n_y,\Tilde{n}_x,\Tilde{n}_y}+\ket{n_x,n_y,\Tilde{n}_x+4,\Tilde{n}_y}\right)+d_2(\ket{n_x,n_y+4,\Tilde{n}_x,\Tilde{n}_y}
\nonumber
\\
&+&\ket{n_x,n_y,\Tilde{n}_x,\Tilde{n}_y+4})+d_3(\ket{n_x+2,n_y+2,\Tilde{n}_x,\Tilde{n}_y}+\ket{n_x,n_y,\Tilde{n}_x+2,\Tilde{n}_y+2})
\nonumber
\\
&+&d_4(\ket{n_x,n_y+2,\Tilde{n}_x+2,\Tilde{n}_y}+\ket{n_x+2,n_y,\Tilde{n}_x,\Tilde{n}_y+2})+(d_5+d_{6})\ket{n_x,n_y,\Tilde{n}_x,\Tilde{n}_y}\Big\}\Big)
\nonumber
\\
&+&\int_0^tdt_2\int_0^tdt_1\gamma^I(t_2)\gamma^I(t_1)\Big\{ a_1(\ket{n_x+2,n_y,\Tilde{n}_x,\Tilde{n}_y+2}+\ket{n_x,n_y+2,\Tilde{n}_x+2,\Tilde{n}_y})
\nonumber
\\
&+&a_2\ket{n_x,n_y+2,\Tilde{n}_x,\Tilde{n}_y+2}+a_3\ket{n_x+2,n_y,\Tilde{n}_x+2,\Tilde{n}_y}+b_1\ket{n_x,n_y+2,\Tilde{n}_x+2,\Tilde{n}_y}
\nonumber
\\
&+&b_2\ket{n_x,n_y,\Tilde{n}_x,\Tilde{n}_y}+b_3\ket{n_x+2,n_y,\Tilde{n}_x,\Tilde{n}_y+2}+c_1(\ket{n_x,n_y,\Tilde{n}_x+4,\Tilde{n}_y}+\ket{n_x+4,n_y,\Tilde{n}_x,\Tilde{n}_y})
\nonumber
\\
&+&c_2(\ket{n_x,n_y,\Tilde{n}_x,\Tilde{n}_y+4}+\ket{n_x,n_y+4,\Tilde{n}_x,\Tilde{n}_y})+c_3(\ket{n_x,n_y,\Tilde{n}_x+2,\Tilde{n}_y+2}+\ket{n_x+2,n_y+2,\Tilde{n}_x,\Tilde{n}_y})
\nonumber
\\
&+&\frac{1}{2}\Big(d_1\left(\ket{n_x+4,n_y,\Tilde{n}_x,\Tilde{n}_y}+\ket{n_x,n_y,\Tilde{n}_x+4,\Tilde{n}_y}\right)+d_2(\ket{n_x,n_y+4,\Tilde{n}_x,\Tilde{n}_y}
\nonumber
\\
&+&\ket{n_x,n_y,\Tilde{n}_x,\Tilde{n}_y+4})+d_3(\ket{n_x+2,n_y+2,\Tilde{n}_x,\Tilde{n}_y}+\ket{n_x,n_y,\Tilde{n}_x+2,\Tilde{n}_y+2})
\nonumber
\\
&+&d_4(\ket{n_x,n_y+2,\Tilde{n}_x+2,\Tilde{n}_y}+\ket{n_x+2,n_y,\Tilde{n}_x,\Tilde{n}_y+2})+(d_5+d_{6})\ket{n_x,n_y,\Tilde{n}_x,\Tilde{n}_y}\Big)\Big\}\Bigg]\otimes\ket{G}
\end{eqnarray}
where $\alpha^{(1)}_{i}$s are defined in \eqref{coeff0}. The other co-efficients are as follows:
\begin{eqnarray}\label{abcd}
        &&d_1=e^{T_+}M^0_{n_x+4,n_x+2}(1)+e^{-T_+}M^2_{n_x+2,n_x+4}(1),
        \nonumber
        \\
        &&d_2=e^{T_+}M^0_{n_y+4,n_y+2}(1)+e^{-T_+}M^2_{n_y+2,n_y+4}(1),
        \nonumber
        \\
        &&d_3=e^{T_+}M^0_{n_x+2,n_y+2}(-2)+e^{-T_+}M^2_{n_x+2,n_y+2}(-2),
        \nonumber
        \\
        &&d_4=2\cos{(T)}M^1_{n_x+2,n_y+2}(2),
        \nonumber
        \\
        &&d_5=e^{-T_-}(M^0_{n_x,n_x}(-2)+M^0_{n_y,n_y}(-2)),
        \nonumber
        \\
        &&d_{6}=e^{T_-}(M^0_{n_x+2,n_x+2}(-2)+M^0_{n_y+2,n_y+2}(-2)),
        \nonumber
        \\
        &&a_1=e^{T_+}M^0_{n_x+2,n_y+2}(1),
        \nonumber
        \\
        &&a_2=e^{T_+}M^0_{n_y+2,n_y+2}(-1), 
        \nonumber
        \\
        &&a_3=e^{T_+}M^0_{n_x+2,n_x+2}(-1),
        \nonumber
        \\
        &&b_1=e^{-T+}M^2_{n_x+2,n_y+2}(1),
        \nonumber
        \\
        &&b_2=e^{-T+}(M^1_{n_x+2,n_x+2}(-1)+M^1_{n_y+2,n_y+2}(-1)), 
        \nonumber
        \\
        &&b_3=e^{-T+}M^2_{n_x+2,n_y+2}(1), 
        \nonumber
        \\
        &&c_1=e^{-\beta\omega}\cos{(T)}M^0_{n_x+4,n_x+2}(1),
        \nonumber
        \\
        &&c_2=e^{-\beta\omega}\cos{(T)}M^0_{n_y+4,n_y+2}(1), 
        \nonumber
        \\
        &&c_3=e^{-\beta\omega}\cos{(T)}M^0_{n_x+2,n_y+2}(-2)~,
\end{eqnarray}
where 
\begin{equation}\label{compact_coeff}
    M^k_{m,n}(s)=se^{-k\beta\omega}\sqrt{m(m-1)}\sqrt{n(n-1)}=Q^k_{m,n}(s)R_{m,n}~.
\end{equation}
In the above we denote $Q^k_{m,n}(s)=se^{-k\beta\omega}\sqrt{mn}$,~ $R_{m,n}=\sqrt{(m-1)(n-1)}$ with $k=0,1,2$, $s=-2,-1,1,2$ and $m,n=n_x+4,n_x+2,n_x,n_y,n_y+2,n_y+4$~.
Also we define $T=2\omega(t_1-t_2)$, $T_+=2i\omega(t_1+t_2)$ and $T_-=2i\omega(t_1-t_2)=iT$.
Expressing \eqref{Int picture g.s} with a more compact notation where we have clubbed the coefficients which contain the information of the graviton-detector system corresponding to each state $\ket{n_x+a,\; n_y+b,\; \tilde{n}_x+c,\; \tilde{n}_y+d}$, one finds
\begin{eqnarray}
        \ket{0,0;G;t}_{\beta}^I&=&(1-e^{-\beta\omega})\sum_{n_x,n_y}e^{-n_x\beta\omega/2}e^{-n_y\beta\omega/2}\Bigg[\alpha^{(1)}_1\ket{n_x+2,n_y,\Tilde{n}_x,\Tilde{n}_y}+\alpha^{(1)}_2\ket{n_x,n_y,\Tilde{n}_x+2,\Tilde{n}_y}
        \nonumber
        \\
&+&\alpha^{(1)}_3\ket{n_x,n_y+2,\Tilde{n}_x,\Tilde{n}_y}+\alpha^{(1)}_4\ket{n_x,n_y,\Tilde{n}_x,\Tilde{n}_y+2}
\nonumber
\\
&+&\left(1+\delta^{(2)}+\epsilon^{(2)}_2+\beta^{(2)}_5+\beta^{(2)}_6+\xi^{(2)}_5+\xi^{(2)}_6\right)\ket{n_x,n_y,\Tilde{n}_x,\Tilde{n}_y}
\nonumber
\\
&+&\left(\beta^{(2)}_1+\lambda^{(2)}_1+\xi^{(2)}_1\right)\left(\ket{n_x+4,n_y,\Tilde{n}_x,\Tilde{n}_y}+\ket{n_x,n_y,\Tilde{n}_x+4,\Tilde{n}_y}\right)
\nonumber
\\
&+&\left(\beta^{(2)}_2+\lambda^{(2)}_2+\xi^{(2)}_2\right)\left(\ket{n_x,n_y+4,\Tilde{n}_x,\Tilde{n}_y}+\ket{n_x,n_y,\Tilde{n}_x,\Tilde{n}_y+4}\right)
\nonumber
\\
&+&\left(\beta^{(2)}_3+\lambda^{(2)}_3+\xi^{(2)}_3\right)\left(\ket{n_x+2,n_y+2,\Tilde{n}_x,\Tilde{n}_y}+\ket{n_x,n_y,\Tilde{n}_x+2,\Tilde{n}_y+2}\right)
\nonumber
\\
&+&\left(\beta^{(2)}_4+\gamma^{(2)}_1+\epsilon^{(2)}_1+\xi^{(2)}_4\right)\ket{n_x,n_y+2,\Tilde{n}_x+2,\Tilde{n}_y}
\nonumber
\\
&+&\left(\beta^{(2)}_4+\gamma^{(2)}_1+\epsilon^{(2)}_3+\xi^{(2)}_4\right)\ket{n_x+2,n_y,\Tilde{n}_x,\Tilde{n}_y+2}
\nonumber
\\
&+&\gamma^{(2)}_2\ket{n_x,n_y+2,\Tilde{n}_x,\Tilde{n}_y+2}+\gamma^{(2)}_3\ket{n_x+2,n_y,\Tilde{n}_x+2,\Tilde{n}_y}\Bigg]\otimes\ket{G}~,
\end{eqnarray}
where $\alpha^{(1)}_i, \delta^{(2)},\beta^{(2)}_j, \gamma^{(2)}_i, \epsilon^{(2)}_k, \lambda^{(2)}_l,\xi^{(2)}_j$ are given as in \eqref{coeff0}.
Introducing a new notation $\hat{C}^{(k)}_{abcd}(t)$ where $k$ indicates the order of perturbation through the number of $\hat{\gamma}(t)$s and $\{a,b,c,d\}$ are the occupation shifts corresponding to $\{n_x,n_y,\Tilde{n}_x,\Tilde{n}_y\}$, we get Eq. \eqref{eq:thermal_state_compact}.

\section{Evaluation of Eq. \eqref{thermal_density_matrix_final_compact}}\label{Appendix D}
Tracing out the unobserved modes, namely $(2,\Tilde{2})$ from $\hat{\rho}^\beta(t)$ (given in Eq. (\ref{rho})), we get the density matrix corresponding to the thermal vacuum state of first HO which is presented below in explicit form: 
\begin{align}\label{thermal density matrix}
    \hat{\rho}^\beta_{1\Tilde{1}}(t)&=\sum_{n_x,n'_x}e^{-(n_x+n'_x)\beta\omega/2}\Bigg[\left(A^{(0)}+B^{(2)}_{n_x,n'_x}\right)\ket{n_x,\Tilde{n}_x}\bra{n'_x,\Tilde{n}'_x}
    +D'^{(1)}_{n_x}\ket{n_x,\Tilde{n}_x+2}\bra{n'_x,\Tilde{n}'_x}\notag\\&+E'^{(1)}_{n_x}\ket{n_x+2,\Tilde{n}_x}\bra{n'_x,\Tilde{n}'_x}
    +D'^{\dagger(1)}_{n'_x}\ket{n_x,\Tilde{n}_x}\bra{n'_x,\Tilde{n}'_x+2}+E'^{\dagger(1)}_{n'_x}\ket{n_x,\Tilde{n}_x}\bra{n'_x+2,\Tilde{n}'_x}\notag\\&+F'^{(2)}_{n_x,n'_x}\left(\ket{n_x,\Tilde{n}_x+2}\bra{n'_x,\Tilde{n}'_x+2}
    +\ket{n_x+2,\Tilde{n}_x}\bra{n'_x+2,\Tilde{n}'_x}\right)\notag\\&+G'^{(2)}_{n_x,n'_x}\left(\ket{n_x+2,\Tilde{n}_x}\bra{n'_x,\Tilde{n}'_x+2}+\ket{n_x,\Tilde{n}_x+2}\bra{n'_x+2,\Tilde{n}'_x}\right)\notag\\&
    +H'^{(2)}_{n_x}\ket{n_x+4,\Tilde{n}_x}\bra{n'_x,\Tilde{n}'_x}+H'^{(2)}_{n_x}\ket{n_x,\Tilde{n}_x+4}\bra{n'_x,\Tilde{n}'_x}\notag\\&+H'^{\dagger(2)}_{n'_x}\ket{n_x,\Tilde{n}_x}\bra{n'_x+4,\Tilde{n}'_x}+H'^{\dagger(2)}_{n'_x}\ket{n_x,\Tilde{n}_x}\bra{n'_x,\Tilde{n}'_x+4}\Bigg]~,
\end{align}
where the coefficients are provided as:
\begin{eqnarray}\label{coeffD}
        &&A^{(0)}=(1-e^{-\beta\omega}),~D'^{(1)}_{n_x}=A^{(0)}\ev{\alpha^{(1)}_2}_G,~E'^{(1)}_{n_x}=A^{(0)}\ev{\alpha^{(1)}_1}_G,
        \nonumber
        \\
        &&F'^{(2)}_{n_x,n'_x}=-G'^{(2)}_{n_x,n'_x}=\left[M^0_{n_x+2,n'_x+2}(1)+M^1_{n'_x+2,n_x+2}(-1)\right]\ev{\abs{\alpha^\beta}^2}_G,
        \nonumber
        \\
        &&H'^{(2)}_{n_x}=\left[M^0_{n_x+4,n_x+2}(1)+M^1_{n_x+2,n_x+4}(-1)\right]\ev{\Gamma^\beta(t)}_G, 
        \nonumber
        \\
        &&B^{(2)}_{n_x,n'_x}=\Big\{Q^0_{n_x,n_x}(1)+Q^1_{n_x,n_x}(-1)+1\Big\}\ev{\delta^{(2)}}_G+\Big\{Q^0_{n'_x,n'_x}(1)+Q^1_{n'_x,n'_x}(-1)+1\Big\}\ev{\delta^{\dagger(2)}}_G
        \nonumber
        \\
        &&+\Big\{M^0_{n_x+2,n_x+2}(-2)+M^1_{n_x+2,n_x+2}(2)\Big\}\ev{\beta^{(2)}_{6}+e^{-\beta\omega}\epsilon^{(2)}_2+\xi^{(2)}_{6}}_G
        \nonumber
        \\
        &&+\Big\{M^0_{n'_x+2,n'_x+2}(-2)+M^1_{n'_x+2,n'_x+2}(2)\Big\}\ev{\beta^{\dagger(2)}_{6}+e^{-\beta\omega}\epsilon^{\dagger(2)}_2+\xi^{\dagger(2)}_{6}}_G
        \nonumber
        \\
        &&+\Big\{M^0_{n_x,n_x}(-2)+M^1_{n_x,n_x}(2)\Big\}\ev{\beta^{(2)}_{5}+\xi^{(2)}_{5}}_G+\Big\{M^0_{n'_x,n'_x}(-2)+M^1_{n'_x,n'_x}(2)\Big\}\ev{\beta^{\dagger(2)}_{5}+\xi^{\dagger(2)}_{5}}_G
        \nonumber
        \\
        &&-2\zeta(\beta)\ev{\beta^{(2)}_{6}+e^{-\beta\omega}\epsilon^{(2)}_2+\xi^{(2)}_{6}+\beta^{\dagger(2)}_{6}+e^{-\beta\omega}\epsilon^{\dagger(2)}_2+\xi^{\dagger(2)}_{6}+\frac{e^{-\beta\omega}}{2}\left(\gamma^{(2)}_2+\gamma^{\dagger(2)}_2\right)}_G
        \nonumber
        \\
        &&+2\zeta(\beta)\ev{\abs{\alpha^\beta}^2-e^{-2\beta\omega}\left(\beta^{(2)}_{5}+\xi^{(2)}_{5}+\beta^{\dagger(2)}_{5}+\xi^{\dagger(2)}_{5}\right)}_G~.
\end{eqnarray}
Again, in Eq. \eqref{coeffD}, we have used \eqref{compact_coeff} and \eqref{coeff0} and introduced $\zeta(\beta)=\frac{2}{(1-e^{-\beta\omega})}$, with $\Gamma^\beta(t)=X^{0~T_+}_{12}(t)+X^{2~-T_+}_{12}(t)+Y^{1~T_-}_{21}(t)+Y^{1~-T_-}_{21}(t)+Y^{0~T_+}_{21}(t)+Y^{2~-T_+}_{21}(t)$. Here we denote
\begin{eqnarray}
&&X^{s\tau}_{ij}(t)=\frac{e^{-s\beta\omega}}{2}\int^t_0 dt_i\int^{t_i}_0 dt_j\comm{\hat{\gamma}^I(t_i)}{\hat{\gamma}^I(t_j)}e^{\tau}~;
\nonumber
\\
&&Y^{s\tau}_{ij}(t)=\frac{e^{-s\beta\omega}}{2}\int^t_0 dt_i\int^{t}_0 dt_j\hat{\gamma}^I(t_i)\hat{\gamma}^I(t_j)e^{\tau}~,
\end{eqnarray}
where it is to be kept in mind that $\tau$ depends on $t_i,t_j$ and also, $i,j=1,2$,~$s=0,2$,~$\tau=\pm T_+,\pm T_-$.
 Further, we have completely extracted the detector depended part from the graviton depended part by redefining $\beta^{(2)}_{j}\rightarrow f(n_x,n'_x,n)\beta^{(2)}_{j} $, $\epsilon^{(2)}_{k}\rightarrow f(n_x,n'_x,n)\epsilon^{(2)}_{k}$, $\xi^{(2)}_{j}\rightarrow f(n_x,n'_x,n)\xi^{(2)}_{j}$ , $\gamma^{(2)}_{i}\rightarrow f(n_x,n'_x,n)\gamma^{(2)}_{i}$ before summing over $n$.
For calculational convenience, clubbing the exponential part to the coefficients in \eqref{thermal density matrix} we first redefine them as: 
\begin{equation*}
    \begin{split}
        A^{(0)}_{n_x,n'_x}=e^{-(n_x+n'_x)\beta\omega/2} A^{(0)},\quad  B^{(2)}_{n_x,n'_x}=e^{-(n_x+n'_x)\beta\omega/2}B^{(2)}_{n_x,n'_x},\quad  
        D^{(1)}_{n_x,n'_x}=e^{-(n_x+n'_x)\beta\omega/2}D'^{(1)}_{n_x},\\ D'^{\dagger(1)}_{n_x,n'_x}=e^{-(n_x+n'_x)\beta\omega/2}D'^{\dagger(1)}_{n'_x},\quad E^{(1)}_{n_x,n'_x}=e^{-(n_x+n'_x)\beta\omega/2}E'^{(1)}_{n_x},\quad E'^{\dagger(1)}_{n_x,n'_x}=e^{-(n_x+n'_x)\beta\omega/2}E'^{\dagger(1)}_{n'_x},\\ F^{(2)}_{n_x,n'_x}=e^{-(n_x+n'_x)\beta\omega/2}F'^{(2)}_{n_x,n'_x},\quad
        G^{(2)}_{n_x,n'_x}=e^{-(n_x+n'_x)\beta\omega/2}G'^{(2)}_{n_x,n'_x},\quad H^{(2)}_{n_x,n'_x}=e^{-(n_x+n'_x)\beta\omega/2}H'^{(2)}_{n_x},\\ H'^{\dagger(2)}_{n_x,n'_x}=e^{-(n_x+n'_x)\beta\omega/2}H'^{\dagger(2)}_{n'_x}~.
    \end{split}
\end{equation*}
Then putting back these redefined coefficients, \eqref{thermal density matrix} can be rewritten as given in \eqref{thermal_density_matrix_final_compact}.
\end{appendix}

\bibliographystyle{apsrev}
\bibliography{bibtexfile1.bib}

\end{document}